\documentclass[11pt,a4paper]{article}
\usepackage{braket}
\usepackage[english]{babel}
\usepackage[utf8]{inputenc}
\usepackage{amsmath}
\usepackage{csquotes}
\usepackage{array, url}
\usepackage{fancyhdr}
\usepackage[top=0.9in,left=1.1in,right=1.1in,bottom=0.9in,headsep=15pt]{geometry}
\usepackage[style=numeric-comp,sorting=none,backend=biber]{biblatex}
\bibliography{defectsv3.bib}
\usepackage{amsfonts}
\usepackage{amsthm}
\usepackage{mathtools}
\usepackage{bm}
\usepackage{titlesec}
\titlespacing*{\section}{0pt}{0.2cm}{0.2cm}
\titlespacing*{\subsection}{0pt}{0.2cm}{0.1cm}
\titlespacing*{\subsubsection}{0pt}{0.3cm}{0.1cm}
\usepackage{titlesec}
\usepackage{graphicx}
\usepackage{epstopdf}
\usepackage{tensor}
\usepackage{comment}
\usepackage{mathrsfs}
\DeclareGraphicsExtensions{.eps}
\usepackage{caption}
\usepackage{subcaption}
\usepackage{etoolbox}
\usepackage{tikz}
\usepackage{authblk}
\usepackage[title]{appendix}
\numberwithin{equation}{section}

\usepackage{amsmath}

\usepackage{ae,aecompl}

\usepackage{hyperref}
\usepackage{xcolor}
\hypersetup{
	colorlinks,
	linkcolor={red!80!black},
	citecolor={blue!80!black},
	urlcolor={blue!80!black},
	linktoc=page
}

\author{\large Jani Kastikainen\thanks{jani.kastikainen@helsinki.fi}}
\affil{\normalsize Department of Physics,\\P.O.Box 64, FIN-00014 \\ University of Helsinki, Finland}
\affil{\normalsize APC, AstroParticule et Cosmologie, Universit\'e Paris Diderot, \\
	CNRS/IN2P3, CEA/IRFU,
	Observatoire de Paris, Sorbonne Paris Cit\'e,\\
	10, rue Alice Domon et L\'eonie Duquet, 75205 Paris Cedex 13, France}
\title{{\Large \textbf{Conical defects and holography\\in topological AdS gravity}}}
\date{}
\begin{document}
	
	\begin{titlepage}
		
		\maketitle
	
		\begin{abstract}
			\noindent We study codimension-even conical defects that contain a deficit solid angle around each point along the defect. We show that they lead to a delta function contribution to the Lovelock scalar and we compute the contribution by two methods. We then show that these codimension-even defects appear as Euclidean brane solutions in higher dimensional topological AdS gravity which is Lovelock--Chern--Simons gravity without torsion. The theory possesses a holographic Weyl anomaly that is purely of type-A and proportional to the Lovelock scalar. Using the formula for the defect contribution, we prove a holographic duality between codimension-even defect partition functions and codimension-even brane on-shell actions in Euclidean signature. More specifically, we find that the logarithmic divergences match, because the Lovelock--Chern--Simons action localizes on the brane exactly. We demonstrate the duality explicitly for a spherical defect on the boundary which extends as a codimension-even hyperbolic brane into the bulk. For vanishing brane tension, the geometry is a foliation of Euclidean AdS space that provides a one-parameter generalization of AdS--Rindler space.
		\end{abstract}
		
	\end{titlepage}

\tableofcontents

\section{Introduction}

Conical singularities have recently played an important role in the context of the AdS\slash CFT correspondence. In particular, they appear when computing conformal field theory (CFT) entanglement entropies of subregions using the replica trick \cite{calabrese_entanglement_2004,calabrese_entanglement_2009}. The replica trick leads to a proof \cite{lewkowycz_generalized_2013} of the Ryu--Takayanagi formula \cite{ryu_holographic_2006,ryu_aspects_2006} that identifies entanglement entropy as the area of a minimal surface in the bulk. The proof also extends to R\'enyi entropy which was shown to be computed by the area of a backreacting cosmic brane \cite{dong_holographic_2014} consisting of conical singularities distributed along a codimension-2 surface.

Geometrically, a cone consists of a compact manifold that shrinks to zero size at the tip of the cone leading to a curvature singularity. The singularity can be point-like or extended along a surface forming a defect or a brane. Usually the manifold that shrinks is a circle so that the defect is of codimension-2, however, more complicated manifolds appear for example in string theory conifolds \cite{candelas_pair_1991,strominger_massless_1995} and the corresponding defect can be of higher codimension. In particular, higher codimension branes have been studied as possible braneworld models in Lovelock gravity \cite{charmousis_matching_2005,appleby_regularized_2007,zegers_self-gravitating_2008} where they appear as solutions.

Lovelock gravities are the most general theories of gravity whose actions depend only on the metric and whose equations of motion are of second order in derivatives \cite{lovelock_einstein_1971}. The Lovelock action is a general linear combination of Lovelock scalars
\begin{equation}
\mathcal{R}_{(m)} = \frac{1}{2^m}\delta^{a_1b_1\ldots a_mb_m}_{c_1d_1\ldots c_md_m}R^{c_1d_1}_{a_1b_1}\cdots R^{c_md_m}_{a_mb_m},
\end{equation}
which are antisymmetrized products of Riemann tensors. Their behaviour in the presence of conical singularities was studied in \cite{fursaev_description_1995}. There it was shown that the contribution of a codimension-2 defect is distributional in nature: the integral over the defect gives a finite contribution and takes the form of the intrinsic Lovelock scalar $ \widehat{\mathcal{R}}_{(m-1)} $ integrated over the defect. The derivation in \cite{fursaev_description_1995} applies to cones with rotational isometry, but it was extended to squashed cones with broken rotational isometry in \cite{fursaev_distributional_2013}.

The strategy used in \cite{fursaev_description_1995,fursaev_distributional_2013} to derive the contribution from a codimension-2 defect is to introduce a small parameter $ \varepsilon $ that smooths out the singularities defining a regular manifold. One then computes the integral of the curvature invariant $ \int \mathcal{R}_{(m)} $ over the regular manifold taking $ \varepsilon\rightarrow 0 $ limit at the end of the computation. The limiting procedure gives rise to an additional finite term proportional to the integral of the intrinsic Lovelock scalar $ \widehat{\mathcal{R}}_{(m-1)} $ of the defect. A second method, applied to the Ricci scalar $ \mathcal{R}_{(1)} $ in \cite{lewkowycz_generalized_2013,dong_gravity_2016}, is to consider the variation $ \delta \int \mathcal{R}_{(1)} $ with respect to the deficit angle of the singularity. In that case, an additional finite term arises from a boundary term localized at the singularity and the result agrees with the regularization method.

In the first part of this work, we use both of the above methods to derive the contribution of a codimension-$ 2p $ defect to $ \int \mathcal{R}_{(m)} $. The defects we consider have a sphere $ S^{2p-1} $ shrinking to zero size at the tip of the cone and contain a solid angle deficit parametrized by a single parameter $ \alpha $. We find that the defect contribution is an integral of the intrinsic Lovelock scalar $ \widehat{\mathcal{R}}_{(m-p)} $ of the defect which generalizes the codimension-2 result. Using the variational method, we also show that the same result arises from the Gibbons--Hawking boundary term of pure Lovelock gravity \cite{teitelboim_dimensionally_1987,myers_higher-derivative_1987}. 

A simple theory of gravity where conical singularities appear as solutions is three-di\-men\-sion\-al Einstein gravity in AdS space \cite{deser_three-dimensional_1984,deser_three-dimensional_1984-1,miskovic_negative_2009}. This theory is a Chern--Simons theory of the AdS isometry group so that all its solutions are locally AdS \cite{deser_three-dimensional_1984-1} and non-trivial effects arise at the global level only. In addition to Einstein gravity, there is a whole family of ($ 2m+1 $)-dimensional Lovelock--Chern--Simons (LCS) gravities in AdS space whose Lagrangians are Chern--Simons forms \cite{chamseddine_topological_1989,chamseddine_topological_1990}. These theories are Lovelock gravity theories with specific values for the couplings that lead to an enhanced local AdS symmetry. The solutions of these theories are not locally AdS, but they are still trivial in the sense of having a flat connection of the corresponding curvature. In this paper, we focus on torsionless (Riemannian) geometries so that the LCS action is a function of the metric only. The solutions are then trivial by being locally Lovelock--AdS metrics.

Three-dimensional Einstein gravity is a well studied toy model for holography and one can ask what aspects of holography in that case generalize to LCS gravity? One aspect is the holographic Weyl anomaly which in LCS gravity is proportional to the Euler characteristic \cite{banados_chern-simons_2004,banados_counterterms_2005,banados_holographic_2006,cvetkovic_holography_2017}. Hence the anomaly is purely of type-A in the classification of \cite{deser_geometric_1993} and, because of the vanishing of the type-B anomaly, the potential dual conformal field theory is necessarily non-unitary \cite{banados_chern-simons_2004}.\footnote{Non-unitary CFTs and holography have been studied for example in \cite{vafa_non-unitary_2014}.} Regardless, one can use the Weyl anomaly to compute partition functions of defects of the potential non-unitary CFT.

In two-dimensional CFTs, defect partition functions compute R\'enyi entropies which are dual to boundary anchored cosmic string solutions \cite{dong_gravity_2016}. In higher dimensions, strings are replaced by branes of which the standard example is the hyperbolic black hole solution \cite{birmingham_topological_1999} that computes Rényi entropy of a ball-shaped region \cite{casini_towards_2011,hung_holographic_2011}. Brane solutions can also be found in LCS gravity which couples to them consistently \cite{miskovic_couplings_2009,edelstein_naked_2011}. It also supports point-particle solutions with a deficit solid angle around the particle that also extend to brane solutions \cite{kastor_black_2006}.\footnote{The point-particle solution in \cite{kastor_black_2006} is for a vanishing cosmological constant, but for the brane solutions, the cosmological constant is non-zero.} In the same vein, we find an Euclidean codimension-$ 2p $ hyperbolic brane solution with a deficit solid angle at each point along the brane.

Codimension-$ 2p $ Euclidean brane solutions that reach the conformal boundary asymptote to codimension-$ 2p $ defects embedded in the conformal boundary geometry. The on-shell brane action of LCS gravity is then a functional of the boundary defect metric whose variation with respect to a Weyl transformation produces the holographic Weyl anomaly $ \mathcal{R}_{(m)} $. The presence of the defect leads to an extra contribution to the anomaly which can be computed using the formula for $ \int \mathcal{R}_{(m)} $ for manifolds with defects. By scale invariance we then obtain the coefficient of the logarithmic divergence in the expansion of the on-shell brane action. The coefficient is simply given by a lower order Lovelock scalar $ \widehat{\mathcal{R}}_{(m-p)} $ of the defect.

Another way to extract the coefficient of the logarithmic divergence is to compute the on-shell brane action directly and, remarkably, we find that the LCS Lagrangian $ \mathcal{L}_{(m)} $ localizes on the brane exactly. In other words, the defect contribution is an integral of the lower dimensional LCS Lagrangian $ \widehat{\mathcal{L}}_{(m-p)} $ of the brane. By expanding the resulting action near the boundary, we find the same coefficient as predicted by the holographic Weyl anomaly of the defect. This is a strong consistency check of our defect formula and of boundary anchored codimension-$ 2p $ branes in LCS gravity.

The simplest setup to demonstrate these computations explicitly is a spherical defect of fixed radius on the boundary. The defect geometry is obtained by transferring to a new set of coordinates via a generalization of the Casini--Huerta--Myers map \cite{casini_towards_2011}: it is a conformal transformation from $ \mathbb{R}^{2m} $ to $ S^{2p-1}\times \mathbb{H}^{2m-2p+1} $. The brane solution that asymptotes to the defect is the codimension-$ 2p $ hyperbolic brane mentioned above. For vanishing brane tension, the solution provides a foliation of Euclidean AdS$ _{2m+1} $ by $ S^{2p-1}\times \mathbb{H}^{2m-2p+1} $-slices and it is the higher dimensional analogue of the Euclidean AdS--Rindler space \cite{parikh_rindler-ads/cft_2012}. As expected, the hyperbolic brane action reproduces the Weyl anomaly of the spherical defect.

The paper is structured as follows. In section \ref{sec:scalarsdef} we derive the contribution of the defect to the integral of a Lovelock scalar. We use two methods: regularization method in subsection \ref{subsec:reg} and variational method in subsection \ref{sec:boundary}. In section \ref{sec:LCS}, we move on to study codimension-$ 2p $ brane solutions in Lovelock--Chern--Simons gravity and prove the correspondence between brane on-shell actions and defect partition functions. In section \ref{sec:sphere}, we demonstrate the duality explicitly for a spherical defect which is dual to a hyperbolic brane in the bulk. Technical details of the derivations can be found in Appendices \ref{app:proof} and \ref{app:chern}. Euclidean AdS$ _{2m+1} $ in codimension-$ 2p $ hyperbolic slicing is presented in Appendix \ref{slicing}.

\subsection{Summary of results}

In the first half of this paper, we derive a formula for the integral of a Lovelock scalar $ \mathcal{R}_{(m)} $ of a manifold $ \mathcal{M}_{\alpha} $ that contains a codimension-$ 2p $ defect $ A $. The type of defects we consider have a solid angle deficit at each point along the defect $ A $. Close to the defect, the metric of $ \mathcal{M}_{\alpha} $ takes the form
\begin{equation}
ds^{2} = \rho^{2}\alpha^{2} d\Omega^{2}_{2p-1} + d\rho^{2} + h_{ij}dx^{i}dx^{j}
\end{equation}
where $ A $ is located at $ \rho = 0 $ and $ h_{ij} $ is the induced metric of $ A $. Locally the metric is $ S^{2p-1}\times A $ and it is spherically symmetric around $ \rho = 0 $ for fixed $ x^{i} $. For $ \alpha\neq 1 $, there is a curvature singularity at $ \rho = 0 $ caused by the solid angle deficit. The formula we prove is
\begin{equation}
\int_{\mathcal{M}_{\alpha}}\sqrt{G}\,\mathcal{R}_{(m)} = D_{(m,p)}(\alpha) +\int_{\mathcal{M}_{\alpha}\,\backslash\, A}\sqrt{G}\, \mathcal{R}_{(m)}
\label{sumdefect}
\end{equation}
where
\begin{equation}
D_{(m,p)}(\alpha) = 
\begin{cases}
C_{(m,p)}\,U_{(p)}(\alpha)\int_{A}\sqrt{h}\,\widehat{\mathcal{R}}_{(m-p)}, \quad &p\leq m\\
0, \quad &p>m
\end{cases}
\end{equation}
is the additive and finite contribution arising from the defect. In other words, the defect gives a delta function contribution to $ \mathcal{R}_{(m)} $. The prefactors here are\,\footnote{Here $ B(x;a,b) = \int_{0}^{x}dt\,t^{a-1}(1-t)^{b-1} $ is the incomplete beta function and $ B(p,1\slash 2) = B(1;p,1\slash 2) $ is the beta function.}
\begin{equation}
C_{(m,p)} = \frac{(4\pi)^p\, m!}{(m-p)!}, \quad U_{(p)}(\alpha)  = \frac{(2p-1)!!}{(2p-2)!!}\int_{\alpha}^{1}du\, \left( 1 - u^2 \right)^{p-1} = \frac{B(1-\alpha^{2};p,1\slash 2)}{B(p,1\slash 2)}
\end{equation}
where the function $ U_{(p)}(\alpha) $ is the regularized beta function that satisfies
\begin{equation}
U_{(1)}(\alpha) = 1-\alpha, \quad U_{(p)}(0) = 1, \quad U_{(p)}(1) = 0.
\end{equation}
The formula \eqref{sumdefect} extends previous results \cite{fursaev_description_1995} of codimension-2 defects to codimension-$ 2p $ ones. For the Euler characteristic, the formula \eqref{sumdefect} also takes a remarkably simple form \eqref{mastereuler}. 

We derive the formula by smoothing out the tip of the cone using a regulator function and taking the sharp limit in the end. Turns out that the sharp limit is independent of the regulator function used. The same formula can also be derived by cutting a hole around the tip in which case $ D_{(m,p)}(\alpha) $ arises from boundary terms.

In the second part of this work, we apply the formula to Euclidean brane solutions $ \mathcal{M}_{\alpha} $ in Lovelock--Chern--Simons gravity in $ (2m+1) $-dimensions. We show that the LCS action localizes on the brane exactly:
\begin{equation}
\int_{\mathcal{M}_{\alpha}}\sqrt{G}\,\mathcal{L}_{(m)} = C_{(m,p)}U_{(p)}(\alpha)\int_{\Sigma}\sqrt{h}\,\widehat{\mathcal{L}}_{(m-p)} +\int_{\mathcal{M}_\alpha \backslash\, \Sigma}\sqrt{G}\,\mathcal{L}_{(m)}
\end{equation}
where $ \mathcal{L}_{(m)} $ is the LCS Lagrangian in $ (2m+1) $-dimensions and $ \widehat{\mathcal{L}}_{(m-p)} $ is the intrinsic LCS Lagrangian of the $ (2m-2p+1) $-dimensional brane $ \Sigma $. Assuming the brane is anchored to the boundary, the result leads to the renormalized on-shell action
\begin{equation}
I_{(m)}^{\text{ren}}[\mathcal{M}_\alpha] = C_{(m,p)}U_{(p)}(\alpha)\int_A \sqrt{\sigma}\,\kappa\ell\, \widehat{\mathcal{R}}^{\text{defect}}_{(m-p)}\,\log{\frac{R}{\epsilon}} + I_{(m)}^{\text{ren}}[\mathcal{M}_\alpha \backslash\, \Sigma]
\end{equation}
where $ \widehat{\mathcal{R}}^{\text{defect}}_{(m-p)} $ is the Lovelock scalar of a codimension-$ 2p $ defect $ A $ (with length scale $ R $) on the boundary to which the brane is anchored, $ \epsilon $ is the UV cut-off, $ \kappa $ is a parameter of the LCS Lagrangian and $ \ell $ is the AdS radius. We prove that the coefficient of the logarithmic divergence can be obtained directly from the boundary Weyl anomaly as well.

Finally, we study an explicit example with a spherical defect $ S^{2m-2p}_{R} $ of radius $ R $ on the conformal boundary. We solve the equations of motion to find the dual geometry $ \mathcal{M}_{\alpha} $ which contains a codimension-$ 2p $ hyperbolic brane $ \Sigma $ that asymptotes to the defect on the boundary $ \partial \Sigma = S^{2m-2p}_{R} $. We show explicitly how the on-shell action of the brane produces the logarithmic divergence in the partition function of the spherical defect as expected by the general analysis.

\section{Lovelock scalars in the presence of codimension-even defects}\label{sec:scalarsdef}

In this section, we derive a formula for the contribution $ D_{(m,p)}(\alpha) $ of a codimension-$ 2p $ defect to the Lovelock scalar $ \mathcal{R}_{(m)} $. We will first introduce $ 2p $-dimensional cones and their regularization after which $ D_{(m,p)}(\alpha) $ is computed by two different methods.

The Lovelock scalar is defined as the contraction of the Lovelock tensor which is \cite{kastor_riemann-lovelock_2012,kastor_conformal_2013}
\begin{equation}
\mathcal{R}^{c_1d_1\ldots c_md_m}_{a_1b_1\ldots a_mb_m(m)} \equiv R^{[c_1d_1}_{[a_1b_1}\cdots R^{c_md_m]}_{a_mb_m]}.
\label{lovelocktensor}
\end{equation}
Contracting the indices we get
\begin{equation}
\mathcal{R}_{(m)} = \frac{1}{2^m}\delta^{a_1b_1\ldots a_mb_m}_{c_1d_1\ldots c_md_m}\mathcal{R}^{c_1d_1\ldots c_md_m}_{a_1b_1\ldots a_mb_m(m)} = \frac{1}{2^m}\delta^{a_1b_1\ldots a_mb_m}_{c_1d_1\ldots c_md_m}R^{c_1d_1}_{a_1b_1}\cdots R^{c_md_m}_{a_mb_m}
\end{equation}
with $ \mathcal{R}_{(0)} \equiv 1 $. Here
\begin{equation}
\delta^{a_1\ldots a_n}_{b_1\ldots b_n} = n!\,\delta^{a_1}_{[b_1}\cdots \delta^{a_n}_{b_n]}
\end{equation}
is the generalized Kronecker delta and in our conventions antisymmetrization contains a factor of $ 1\slash n! $. Lovelock scalars form the basis of Lovelock theories of gravity that are the most general actions constructed out of the metric tensor whose equations of motion are second order in the metric (see \cite{padmanabhan_lanczos-lovelock_2013} for a review).

\subsection{Conical defects of even codimension}\label{sec:defect}

\begin{figure}[t]
	\centering
	\includegraphics{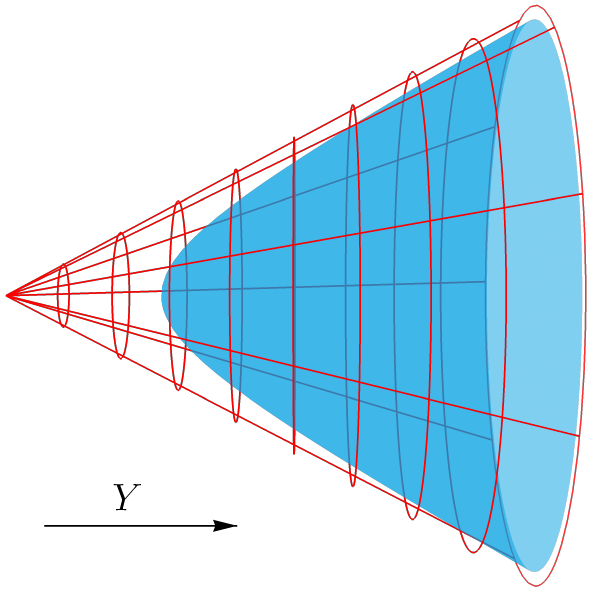}
	\caption{A two-dimensional cone regularized by the function $ f_{\varepsilon}(\rho) = \sqrt{\rho^{2} + \varepsilon^{2}} $ which corresponds to $ u_\varepsilon(\rho) = \frac{\rho^{2}+\alpha^{2}\varepsilon^{2}}{\rho^{2} + \varepsilon^{2}} $.}
	\label{regcone}
\end{figure}

A $ 2p $-dimensional cone (with $ p=1,2,\ldots $) is the surface
\begin{equation}
Y^2- \frac{1-\alpha^2}{\alpha^2}\sum_{i=1}^{2p} X_i^2 = 0, \quad Y \geq 0
\label{cone}
\end{equation}
embedded in $ \mathbb{R}^{2p+1} $ with cartesian coordinates $ (Y,X_i) $. The parameter $ 0<\alpha \leq 1 $ controls the steepness of the cone with $ \alpha = 1 $ being the flat plane $ \mathbb{R}^{2p} $ at $ Y=0 $. It is related the opening angle $ \theta_0\in[0,\pi] $ of the cone as $ \alpha = \sin{(\theta_0\slash 2)} $. The equation of the cone is solved by
\begin{align}
X_i &= \alpha \rho\,\Omega_i\label{emb}\\
Y &= \rho\,\sqrt{1-\alpha^2}
\end{align}
where $ \sum_{i=1}^{2p} \Omega_i^2 = 1 $ parametrize a sphere and $ \rho \geq 0 $ is its radial size. The resulting induced metric of the cone is
\begin{equation}
ds^2 = \rho^2 \alpha^2d\Omega_{2p-1}^2 + d\rho^2.
\label{pcone}
\end{equation}
The case $ p=1 $ corresponds to a two-dimensional cone with a deficit angle $ 2\pi (1-\alpha) $ \cite{fursaev_description_1995} and for $ p>1 $ there is a deficit in the solid angle. In \cite{kastor_black_2006}, the metric \eqref{pcone} describes a point-mass solution of pure Lovelock gravity in the critical dimension. Higher dimensional cones have also been studied in the context of holographic entanglement entropy in \cite{myers_entanglement_2012,bueno_universal_2015}.

Unlike a two-dimensional cone, a $ 2p $-dimensional cone \eqref{pcone} is not flat. Instead, for $ m\geq p $ it is Lovelock flat:
\begin{equation}
\mathcal{R}^{c_1d_1\ldots c_md_m}_{a_1b_1\ldots a_mb_m(m)} = 0, \quad m\geq p
\label{lovelockflatcone}
\end{equation}
which is due to the fact that \eqref{pcone} has only $ p-1 $ non-zero Riemann tensors given by
\begin{equation}
R^{\phi_1\phi_2}_{\varphi_1\varphi_2} = \frac{1-\alpha^2}{\alpha^2\rho^2}\delta^{\phi_1\phi_2}_{\varphi_1\varphi_2}
\label{pointriemann}
\end{equation}
so that the anti-symmetrization in \eqref{lovelockflatcone} vanishes. This means that curvature scalars constructed from the Lovelock tensor are blind to the deficit solid angle parameter $ \alpha $. In \eqref{pointriemann} the indices $ \phi,\varphi, \ldots $ denote the $ 2p-1 $ angular components of $ d\Omega_{2p-1}^2 $.

If $ \alpha < 1 $ the cone contains a singularity at $ \rho = 0 $ where the sphere shrinks to zero size. This can be seen from the embedding function $ Y(\rho) $ which goes to zero with slope $ \sqrt{1-\alpha^2} $ and does not have vanishing derivative at $ \rho = 0 $. It is also evident from the non-zero components of the Riemann tensor \eqref{pointriemann} that blow up at $ \rho = 0 $. In the case $ p=1 $, it is known that the singularity is distributional: it leads to a delta function contribution to $ \mathcal{R}_{(m)} $ which gives a finite contribution inside integrals \cite{fursaev_description_1995}. The same turns out to be true for singularities with $ p>1 $ as we will show.

The sharp tip of the cone can be smoothed out by introducing a regulating function $ f_\varepsilon(\rho) $ that contains an extra length scale $ \varepsilon $ and that satisfies
\begin{equation}
f_0(\rho) = \rho, \quad f'_\varepsilon(0) = 0.
\label{fconds}
\end{equation}
The regularized cone is then the surface \eqref{emb}, but with 
\begin{equation}
Y = f_\varepsilon(\rho)\sqrt{1-\alpha^2}
\end{equation}
so that the slope smoothly goes to zero $ Y'(0) = 0 $ at the tip of the cone (see figure \ref{regcone}). The sharp cone is obtained from the regularized cone in the $ \varepsilon \rightarrow 0 $ limit.

The metric of the regularized cone is given by
\begin{equation}
ds^2 = \rho^2 \alpha^2d\Omega_{2p-1}^2 + u_\varepsilon(\rho)d\rho^2
\label{ucone}
\end{equation}
where $ u_\varepsilon(\rho) $ is related to $ f_\varepsilon(\rho) $ via \cite{fursaev_description_1995}
\begin{equation}
u_\varepsilon(\rho) = f'_\varepsilon(\rho)^2(1-\alpha^2) + \alpha^2.
\label{defu}
\end{equation}
The function $ u_\varepsilon $ has to be dimensionless so by dimensional analysis
\begin{equation}
u_\varepsilon(\rho) = u(\rho\slash \varepsilon).
\end{equation}
Then \eqref{fconds} and \eqref{defu} imply
\begin{equation}
u(s) = \alpha^2 + \ddot{u}(0) s^2 + \mathcal{O}(s^3), \quad \lim_{s \rightarrow \infty} u(s) = 1
\label{boundconds}
\end{equation}
where an overdot denotes a derivative with respect to $ s = \rho \slash \varepsilon $. One can check that the metric \eqref{ucone} is indeed regular by computing the Riemann tensors which in the coordinate $ s=\rho \slash \varepsilon $ are
\begin{equation}
R^{\phi_1\phi_2}_{\varphi_1\varphi_2} = \frac{1}{\varepsilon^2}\frac{u(s) - \alpha^2}{\alpha^2s^2u(s)}\delta^{\phi_1\phi_2}_{\varphi_1\varphi_2}, \quad 
R^{s\phi}_{s\varphi} = \frac{1}{\varepsilon^2}\frac{\dot{u}(s)}{2s u(s)^2}\delta^\phi_\varphi.
\label{riemanns}
\end{equation}
From the boundary conditions \eqref{boundconds} it follows that these are indeed finite at $ s=0 $ when $ \varepsilon \neq 0 $.

We are interested in codimension-$ 2p $ conical defects that have a singularity of the form \eqref{pcone} at each point along an extended surface $ A $. The defect $ A $ is embedded in a $ D $-dimensional Euclidean manifold $ \mathcal{M}_{\alpha} $ and we assume that $ D\geq 2p $ so that the dimension of $ A $ can be zero. The metric $ G_{(0)} $ of $ \mathcal{M}_{\alpha} $ close to the defect then takes the general form
\begin{equation}
ds^2 = f(\rho,\Omega,x)  d\Omega^2_{2p-1} + d\rho^2  + F_{ij}(\rho,\Omega,x)dx^idx^j
\label{extnear}
\end{equation}
where the functions have the expansions
\begin{equation}
f(\rho,\Omega,x) = \rho^2 \alpha^2 + \mathcal{O}(\rho^4), \quad F_{ij}(\rho,\Omega,x) = h_{ij}(x) + \mathcal{O}(\rho^2).
\end{equation}
In these coordinates, the defect $ A $ is located at $ \rho=0 $ and the $ D-2p $ internal dimensions of $ A $ are parametrized by coordinates $ x^i $ with $ h_{ij} $ being its induced metric. We do not impose any additional constraints on the shape of the manifold $ \mathcal{M}_{\alpha} $ outside of the defect.

The metric \eqref{extnear} can be regularized by introducing a regulator $ u(s) $ as above. The regularized near defect metric is denoted by $ G_{\varepsilon(0)} $ and is explicitly
\begin{equation}
ds^2 = \rho^2 \alpha^2 d\Omega^2_{2p-1} +  u(\rho\slash \varepsilon)d\rho^2  + h_{ij}(x)dx^idx^j.
\label{near}
\end{equation}
The transverse (respect to $ A $) Riemann tensors of this metric are given by \eqref{riemanns}.

\subsection{Contribution of the defect to the Lovelock scalar}\label{subsec:reg}

In this section, we compute the finite contribution $ D_{(m,p)}(\alpha) $ to the integral of a Lovelock scalar using the regularization method. The setup is a Euclidean manifold $ \mathcal{M}_{\alpha} $ of dimension $ D\geq 2p $ that contains a codimension-$ 2p $ defect $ A $. The defect is regularized by introducing a parameter $ \varepsilon $ as above. This defines a regular manifold $ \mathcal{M}_\alpha(\varepsilon) $ for which we can calculate the integral of the Lovelock scalar without any problems. Then we take the limit $ \varepsilon\rightarrow 0 $ and extract the extra contribution $ D_{(m,p)}(\alpha) $ coming from the defect. The same strategy was used in \cite{fursaev_description_1995,fursaev_distributional_2013} to compute the contribution of codimension-2 defects. An alternative approach to computing $ D_{(m,p)}(\alpha) $ is presented in section \ref{sec:boundary}.

Denote the metric of $ \mathcal{M}_\alpha(\varepsilon) $ by $ G_{\varepsilon} $ and work in coordinates where the near defect metric is \eqref{near}. As in \cite{fursaev_distributional_2013}, we divide the integral into two pieces around the defect as
\begin{equation}
\int_{\mathcal{M}_\alpha(\varepsilon)}\sqrt{G_\varepsilon}\, \mathcal{R}_{(m)}[G_\varepsilon] =  \int_{\rho \leq \rho_0}\sqrt{G_\varepsilon}\, \mathcal{R}_{(m)}[G_\varepsilon] + \int_{\rho > \rho_0}\sqrt{G_\varepsilon}\, \mathcal{R}_{(m)}[G_\varepsilon]
\label{integral}
\end{equation}
where $ \rho_0 > 0 $ is a radius which will be kept fixed during the $ \varepsilon \rightarrow 0 $ limit. When $ \varepsilon \rightarrow 0 $, the first term integrates over the singularity and will produce $ D_{(m,p)}(\alpha) $. Assuming $ \rho_0 $ is sufficiently small, the first term can be computed using the near defect metric $ G_{\varepsilon(0)} $ \eqref{near} up to corrections that vanish once $ \rho_0 \rightarrow 0 $ is taken. The second term, on the other hand, is regular and we can set $ \varepsilon = 0 $. We get
\begin{equation}
\int_{\mathcal{M}_\alpha}\sqrt{G}\, \mathcal{R}_{(m)}[G] = D_{(m,p)}(\alpha) + \int_{\mathcal{M}_\alpha\backslash \, A}\sqrt{G}\, \mathcal{R}_{(m)}[G]
\end{equation}
where $ G $ is the metric of $ \mathcal{M}_\alpha(0) $. Here
\begin{equation}
D_{(m,p)}(\alpha) = \lim_{\rho_0\rightarrow 0}\lim_{\varepsilon\rightarrow 0}\int_{\rho \leq \rho_0}\sqrt{G_{\varepsilon(0)}}\, \mathcal{R}_{(m)}[G_{\varepsilon(0)}]
\label{Dmp}
\end{equation}
is the contribution from the singularity and
\begin{equation}
\int_{\mathcal{M}_\alpha\backslash \, A}\sqrt{G}\, \mathcal{R}_{(m)}[G] \equiv \lim_{\rho_0 \rightarrow 0}\int_{\rho \geq \rho_0}\sqrt{G}\, \mathcal{R}_{(m)}[G]
\label{regular}
\end{equation}
is the integral over the regular part of the manifold. In other words, \eqref{regular} is the integral from $ \rho=0 $ using the regular part of the metric. It is finite, because the volume form compensates for the diverging curvatures \eqref{pointriemann} as $ \rho\rightarrow 0 $.\footnote{The integrand contains at most $ p-1 $ curvatures \eqref{pointriemann} and goes as $ \rho^{2p-1}(1 \slash \rho^{2(p-1)}) \sim \rho $ as $ \rho \rightarrow 0 $.}

Rest of the section is devoted to the computation of \eqref{Dmp}. We assume that $ p\leq m $ and the case $ p>m $ is handled separately in the end. 

The Lovelock scalar in \eqref{Dmp} consists of a product of $ m $ Riemann tensors summed over the indices $ (s,\Omega,x) $. We can divide the sum into parts depending on the number of $ (s,\Omega) $-components in each one. Due to total anti-symmetrization imposed by the Kronecker delta, each upper or lower index appears only once. It takes the form
\begin{equation}
\mathcal{R}_{(m)}[G_{\varepsilon(0)}] = \sum_{n} \mathcal{R}_{(m,n)}
\label{sumriem}
\end{equation}
where schematically
\begin{equation}
\mathcal{R}_{(m,n)} \cong \overbrace{R^{s\varphi_1}_{s\phi_1}R^{\varphi_2\varphi_3}_{\phi_2\phi_3}\cdots R^{\varphi_{2n}\varphi_{2n+1}}_{\phi_{2n}\phi_{2n+1}}}^{n}\;\overbrace{R^{j_{1}j_{2}}_{i_{1}i_{2}}\cdots R^{j_{2m-2n-1}j_{2m-2n}}_{i_{2m-2n-1}i_{2m-2n}}}^{m-n}
\label{schem}
\end{equation}
with all the indices contracted by the generalized Kronecker delta. Each term is weighted by a combinatorial factor that arises from permuting the indices into order presented.

The upper limit of the sum \eqref{sumriem} is $ n=p $ which is the maximum number of Riemann tensors $ R^{s\varphi}_{s\phi},R^{\varphi_1\varphi_2}_{\phi_1\phi_2} $ \eqref{riemanns} available. Note that if $ p>m $ then the upper limit is $ m $ and not all angular tensors $ R^{\varphi_1\varphi_2}_{\phi_1\phi_2} $ fit into the product. The lower limit, on the other hand, depends on the amount of tangential Riemann tensors $ R^{ij}_{kl} $ available. 

To compute \eqref{Dmp}, we factorize the integration measure as\,\footnote{Here $ \Omega_{2p-1} = \frac{2\pi^p}{(p-1)!} $ is the volume of $ S^{2p-1} $.}
\begin{equation}
\int_{\rho \leq \rho_0}\sqrt{G_{\varepsilon(0)}} = \Omega_{2p-1}\alpha^{2p-1}\int_0^{\rho_0} d\rho\, \rho^{2p-1}\sqrt{u(\rho\slash \varepsilon)} \int_A d^{D-2p}x\,\sqrt{h} 
\end{equation}
where we performed the angular integrals using spherical symmetry. Performing a change of variables $ s = \rho\slash \varepsilon $, we get
\begin{equation}
D_{(m,p)}(\alpha) = \Omega_{2p-1}\sum_n\int_{A}d^{D-2p}x \sqrt{h} \,d_{(m,n)}(x)
\label{rhozero}
\end{equation}
where
\begin{equation}
d_{(m,n)}(x) = \lim_{\rho_0\rightarrow 0}\lim_{\varepsilon \rightarrow 0}\int_0^{\rho_0 \slash \varepsilon} ds\, s^{2p-1}\sqrt{u(s)}\,\alpha^{2p-1}\varepsilon^{2p}\,\mathcal{R}_{(m,n)}
\label{dintegral}
\end{equation}
and the Riemann tensors are written in the coordinate $ s $ \eqref{riemanns}. Each Riemann tensor of the cone \eqref{riemanns} comes with a factor of $ \varepsilon^{-2} $ so that the integrals \eqref{dintegral} have the form
\begin{equation}
d_{(m,n)}(x)\propto \lim_{\rho_0\rightarrow 0}\lim_{\varepsilon\rightarrow 0}\int_0^{\rho_0 \slash \varepsilon} ds\, \varepsilon^{2(p-n)} s^{2p-1} f(s)
\label{propto}
\end{equation}
where $ f(s) $ contains all the $ s $-dependence coming from the Riemann tensors and all the $ \varepsilon $-dependence is included in $ \varepsilon^{2(p-n)} $.

One can see that $ d_{(m,n)} $ with $ n = p $ is special: it does not have any $ \varepsilon $-dependence in the integrand. Hence taking the limit $ \varepsilon \rightarrow 0 $ simply sets the upper limit of the integral to infinity. This gets rid of all the $ \rho_{0} $-dependence as well so that the second limit $ \rho_{0}\rightarrow 0 $ is trivial. For $ d_{(m,n)} $ with $ n<p $ we have to do more work: they vanish in the $ \rho_0,\varepsilon \rightarrow 0 $ limit which is shown in Appendix \ref{app:proof}. As a result, we get
\begin{equation}
D_{(m,p)}(\alpha) = \Omega_{2p-1}\int_{A}d^{D-2p}x \sqrt{h} \,d_{(m,p)}(x)
\label{defectcontr}
\end{equation}
which we will now compute.

\subsubsection*{The non-zero $ n=p $ contribution}

We have explicitly
\begin{align}
\mathcal{R}_{(m,p)} = \frac{4m}{2^m}\binom{m-1}{p-1}\delta^{s\phi_1\phi_2\ldots \phi_{2p-1}i_{1}\ldots i_{2m-2p}}_{s\varphi_1\varphi_2\ldots \varphi_{2p-1}j_{1}\ldots j_{2m-2p}}&\overbrace{R^{s\varphi_1}_{s\phi_1}R^{\varphi_2\varphi_3}_{\phi_2\phi_3}\cdots R^{\varphi_{2p-2}\varphi_{2p-1}}_{\phi_{2p-2}\phi_{2p-1}}}^{p} \nonumber\\
&\times \underbrace{R^{j_{1}j_{2}}_{i_{1}i_{2}}\cdots R^{j_{2m-2p-1}j_{2m-2p}}_{i_{2m-2p-1}i_{2m-2p}}}_{m-p}.
\label{expand}
\end{align}
The degeneracy factor is determined combinatorially as follows. First we have to pick the Riemann tensor that contains the $ s $-indices. There are $ m $ choices each with four ways of arranging the indices in $ R^{s\phi}_{s\varphi} $ due to the symmetries of the Riemann tensor. This gives the factor of $ 4m $. From the remaining $ m-1 $ Riemann tensors we have to pick $ p-1 $ that contain the $ 2(p-1) $ angular components as $ R^{\phi_1\phi_2}_{\varphi_1\varphi_2} $. The order of these in \eqref{expand} does not matter as the sum is invariant under exchange of the tensors yielding the factor of $ \binom{m-1}{p-1} $.

Next note that the Kronecker delta factorizes
\begin{equation}
\delta^{\phi_1\phi_2\ldots \phi_{2p-1}i_{1}\ldots i_{2m-2p}}_{\varphi_1\varphi_2\ldots \varphi_{2p-1}j_{1}\ldots j_{2m-2p}} = \delta^{\phi_1\ldots \phi_{2p-1}}_{\varphi_1\ldots \varphi_{2p-1}}\delta^{i_{1}\ldots i_{2m-2p}}_{j_{1}\ldots j_{2m-2p}}
\end{equation}
so that we get
\begin{align}
\mathcal{R}_{(m,p)} = \frac{4m}{2^{m}}\binom{m-1}{p-1}&\delta^{\phi_1\ldots \phi_{2p-1}}_{\varphi_1\ldots \varphi_{2p-1}}R^{s\varphi_1}_{s\phi_1}R^{\varphi_2\varphi_3}_{\phi_2\phi_3}\cdots R^{\varphi_{2p-2}\varphi_{2p-1}}_{\phi_{2p-2}\phi_{2p-1}} \nonumber\\
&\times\delta^{i_{1}\ldots i_{2m-2p}}_{j_{1}\ldots j_{2m-2p}}R^{j_{1}j_{2}}_{i_{1}i_{2}}\cdots R^{j_{2m-2p-1}j_{2m-2p}}_{i_{2m-2p-1}i_{2m-2p}}.
\label{univ2}
\end{align}
Substituting the angular Riemann tensors \eqref{riemanns} and performing the Kronecker contractions using
\begin{equation}
\delta^{\phi_1\phi_2\ldots \phi_{2p-1}}_{\varphi_1\varphi_2\ldots \varphi_{2p-1}}\delta^{\phi_1}_{\varphi_1}\delta^{\phi_2\phi_3}_{\varphi_2\varphi_3}\cdots \delta^{\phi_{2p-2}\phi_{2p-1}}_{\varphi_{2p-2}\varphi_{2p-1}} = 2^{p-1}(2p-1)!\,,
\end{equation}
we get
\begin{align}
\mathcal{R}_{(m,p)}=2m(2p-1)!\binom{m-1}{p-1}&\frac{1}{2\alpha^{2p-2}}\frac{\dot{u}(s)\left( u(s) - \alpha^2\right)^{p-1}}{s^{2p-1}u(s)^{p+1}}\nonumber\\
&\times\frac{1}{2^{m-p}}\delta^{i_{1}\ldots i_{2m-2p}}_{j_{1}\ldots j_{2m-2p}}R^{j_{1}j_{2}}_{i_{1}i_{2}}\cdots R^{j_{2m-2p-1}j_{2m-2p}}_{i_{2m-2p-1}i_{2m-2p}}.
\end{align}
The near defect metric has spherical symmetry around the defect $ s = 0 $.\footnote{Codimension-2 cones with broken spherical symmetry (squashed cones) were studied in \cite{fursaev_distributional_2013}. Generalization to codimension-$ 2p $ squashed cones is left for future work.} This means that all the $ 2p $ extrinsic curvatures of the $ s = 0 $ surface vanish. Using the Gauss-Codazzi equation, we can hence replace $ R^{ij}_{kl} $ by the intrinsic curvatures $ \widehat{R}^{ij}_{kl} $ of the defect $ A $. The corresponding sum over the latin indices in \eqref{expand} is thus simply the intrinsic Lovelock scalar $ \widehat{\mathcal{R}}_{(m-p)} $. We are left with
\begin{equation}
\mathcal{R}_{(m,p)} = \widetilde{C}_{(m,p)}\frac{1}{\varepsilon^{2p}}\frac{1}{2\alpha^{2p-2}}\frac{\dot{u}(s)\left( u(s) - \alpha^2\right)^{p-1}}{s^{2p-1}u(s)^{p+1}}\,\widehat{\mathcal{R}}_{(m-p)}(x)
\end{equation}
where we have defined\,\footnote{We have checked this combinatorial factor numerically for $ m=2,p=2 $ and $ m=3,p=2 $ finding agreement.}
\begin{equation}
\widetilde{C}_{(m,p)} = 2m(2p-1)!\binom{m-1}{p-1}.
\label{comb}
\end{equation}
Hence the integral \eqref{dintegral} for $ n=p $ becomes
\begin{equation}
d_{(m,p)}(x) = \widetilde{C}_{(m,p)}\widehat{\mathcal{R}}_{(m-p)}(x)\,\frac{\alpha}{2}\int_0^{\infty} ds\, \frac{\dot{u}(s)\left( u(s) - \alpha^2\right)^{p-1}}{u(s)^{p+1\slash 2}}. 
\label{dpminus1}
\end{equation}
where we sent $ \varepsilon \rightarrow 0 $ in the upper limit of the integral as all the $ \varepsilon $-dependence of the integrand cancelled. We can write the $ s $-integral in \eqref{dpminus1} as an $ u $-integral using $ u(0) = \alpha^2 $ and $ u(\infty) = 1 $:
\begin{equation}
\frac{\alpha}{2}\int_{\alpha^2}^{1}du\, \frac{\left( u- \alpha^2 \right)^{p-1}}{u^{p+1\slash 2}} = \int_{\alpha}^{1}du\, \left( 1 - u^2 \right)^{p-1} \equiv \widetilde{U}_{(p)}(\alpha)
\label{singint}
\end{equation}
where the second equality follows by doing a change of variables $ u \rightarrow \alpha \slash \sqrt{u} $ and using the boundary conditions \eqref{boundconds}. This integral is universal in the space of regulator functions: it does not depend on the explicit form of $ u(s) $ and is completely determined by the boundary conditions \eqref{boundconds}. Thus we finally get
\begin{equation}
d_{(m,p)}(x) = \widetilde{C}_{(m,p)}\widetilde{U}_{(p)}(\alpha)\widehat{\mathcal{R}}_{(m-p)}(x).
\label{dmp}
\end{equation}

\subsubsection*{The defect contribution}

Substituting \eqref{dmp} to \eqref{defectcontr}, we get
\begin{equation}
D_{(m,p)}(\alpha) = \Omega_{2p-1}\widetilde{C}_{(m,p)}\widetilde{U}_{(p)}(\alpha)\int_{A}d^{D-2p}x \sqrt{h} \,\widehat{\mathcal{R}}_{(m-p)}
\label{tildes}
\end{equation}
which holds for $ p\leq m $. For $ p > m $, the term $ d_{(m,p)} $ will not be special anymore in the sense explained above. Instead, it will vanish in the same way as the terms with $ n<p $ which is shown in Appendix \ref{app:proof}. Hence
\begin{equation}
D_{(m,p)}(\alpha) = 0, \quad \text{for} \quad p > m.
\end{equation}
Heuristically the vanishing occurs, because the singularity is not strong enough to compensate for the volume form in dimensions $ D\geq 2p $. It is confirmed by the alternative method of computing $ D_{(m,p)}(\alpha) $ in section \ref{sec:boundary} and Appendix \ref{app:chern}. The vanishing is also of fundamental importance when we compute the brane contribution to the Lovelock--Chern--Simons action in section \ref{sec:LCS}.

We will now normalize the integral $ \widetilde{U}_{(p)}(\alpha) $ in the formula \eqref{tildes}. At $ \alpha = 0 $, it is
\begin{equation}
\widetilde{U}_{(p)}(0) = \int^{1}_{0}du\, \left( 1 - u^2 \right)^{p-1} = \frac{4^{p-1}\,(p-1)!^2}{(2p-1)!} = \frac{(2p-2)!!}{(2p-1)!!}.
\label{Uintegral}
\end{equation}
We define the integral
\begin{equation}
U_{(p)}(\alpha)  = \frac{(2p-1)!!}{(2p-2)!!}\int_{\alpha}^{1}du\, \left( 1 - u^2 \right)^{p-1} \quad \Rightarrow \quad U_{(p)}(0) = 1
\end{equation}
which is normalized to unity at $ \alpha=0 $. Noting that
\begin{equation}
\Omega_{2p-1}\widetilde{C}_{(m,p)} = \frac{(4\pi)^p\, m!}{(m-p)!}\frac{(2p-1)!!}{(2p-2)!!}
\end{equation}
we get the final formula
\begin{equation}
D_{(m,p)}(\alpha) =
\begin{cases}
C_{(m,p)}\,U_{(p)}(\alpha)\int_{A}d^{D-2p}x\sqrt{h}\,\widehat{\mathcal{R}}_{(m-p)}, \quad &p\leq m\\
0, \quad &p>m
\end{cases}
\label{master}
\end{equation}
where
\begin{equation}
C_{(m,p)} = \frac{(4\pi)^p\, m!}{(m-p)!}.
\label{combcoeff}
\end{equation}
The function $ U_{(p)}(\alpha) $ can be expressed as the regularized beta function
\begin{equation}
U_{(p)}(\alpha) = I(1-\alpha^{2};p,1\slash 2) = \frac{B(1-\alpha^{2};p,1\slash 2)}{B(p,1\slash 2)}
\end{equation}
where $ B(x;a,b) = \int_{0}^{x}dt\,t^{a-1}(1-t)^{b-1} $ is the incomplete beta function and $ B(p,1\slash 2) = B(1;p,1\slash 2) $ is the beta function.

Equation \eqref{master} contains the two-dimensional conical singularity $ p=1 $ as a special case. Noting that
\begin{equation}
U_{(1)}(\alpha) = 1-\alpha
\end{equation}
gives
\begin{equation}
D_{(m,1)}(\alpha) = 4\pi m (1-\alpha)\int_{A}d^{D-2}x\,\sqrt{h}\,\widehat{\mathcal{R}}_{(m-1)}
\end{equation}
which agrees with \cite{fursaev_description_1995,fursaev_distributional_2013}. Another interesting special case is $ p = m $ for which the dependence on the intrinsic curvature of $ A $ completely disappears from \eqref{master}:
\begin{equation}
D_{(m,m)}(\alpha) = (4\pi)^m\,m!\,U_{(m)}(\alpha)\int_{A}d^{D-2m}x\,\sqrt{h}
\end{equation}
which is proportional to the area of $ A $.

\subsubsection{Formula for the Euler characteristic}

In dimension $ D=2m \geq 2p $, the integral over the Lovelock scalar computes the Euler characteristic $ \chi_{m}[\mathcal{M}] $ of the manifold. For a manifold without boundaries it is defined as (see for example \cite{charmousis_matching_2005} and references therein)
\begin{equation}
\chi_{m}[\mathcal{M}] = \frac{1}{(4\pi)^m\,m!}\int_{\mathcal{M}}d^{2m}x\,\sqrt{G}\,\mathcal{R}_{(m)}
\end{equation}
so that using \eqref{master} for a manifold $ \mathcal{M}_\alpha $ with a conical defect $ A $, we get
\begin{equation}
\chi_{m}[\mathcal{M}_\alpha] = U_{(p)}(\alpha)\,\chi_{m-p}[A] + \chi_{m}[\mathcal{M}_\alpha \backslash A].
\label{mastereuler}
\end{equation}
Remarkably, the prefactors have combined in such a way to yield the Euler characteristic of the defect $ \chi_{m-p}[A] $ multiplied by the normalized function $ U_{(p)}(\alpha) $.

\subsection{Defect contribution from boundary terms}\label{sec:boundary}

An alternative way to derive the contribution of codimension-2 defects to the Ricci scalar was used in \cite{lewkowycz_generalized_2013,dong_gravity_2016}. The idea is to compute the metric variation of $ \int \mathcal{R}_{(1)} $ with respect to a small change in $ \alpha $ so that $ D_{(1,1)}(\alpha) $ arises from a boundary term at $ A $. We will now generalize this approach to codimension-$ 2p $ defects and Lovelock scalars and use it to obtain a formula for $ D_{(m,p)}(\alpha) $ in terms of the Chern form $ B_{(m)} $. In Appendix \ref{app:chern}, we use this formula to verify \eqref{master} derived using the regularization method.

Let $ \mathcal{D}_\epsilon $ be a small tube $ \rho \leq \epsilon $ surrounding the defect $ A $ in the geometry $ \mathcal{M}_\alpha $. Denote the metric on the tube boundary $ \partial \mathcal{D}_\epsilon $ ($ \rho = \epsilon $) by $ H_{\mu\nu} $ with the indices $ \mu,\nu $ running over the coordinates $ (\Omega,x) $. Then consider the manifold $ \mathcal{M}_\alpha\,\backslash \, \mathcal{D}_\epsilon $ with the tube removed and perform a metric variation $ \delta_{\alpha} $ that varies $ \alpha $. Then \cite{chakraborty_novel_2017}
\begin{equation}
\delta_\alpha \int_{\mathcal{M}_\alpha\,\backslash \, \mathcal{D}_\epsilon}\sqrt{G}\,\mathcal{R}_{(m)} =  -\delta_\alpha\int_{\partial \mathcal{D}_\epsilon}\sqrt{H}\,B_{(m)}+\int_{\partial \mathcal{D}_\epsilon}\sqrt{H}\,\tau^{\mu\nu}_{(m)}\delta_\alpha H_{\mu \nu} + \int_{\mathcal{M}_\alpha\,\backslash \, \mathcal{D}_\epsilon} \sqrt{G}\,E_{b(m)}^a\delta_\alpha G^{b}_a
\label{riemder}
\end{equation}
where $ \tau_{\mu\nu(m)} $ is the boundary stress-energy tensor of pure Lovelock gravity, $ E_{ab(m)} $ is the corresponding equation of motion tensor and $ B_{(m)} $ is the Chern form \cite{chern_curvatura_1945}:
\begin{equation}
B_{(m)} = 2m\int_0^1dt\,\delta^{\mu\mu_1\nu_1\ldots \mu_{m-1}\nu_{m-1}}_{\nu\rho_1\sigma_1\ldots \rho_{m-1}\sigma_{m-1}}K^{\nu}_{\mu}\prod_{k=1}^{m-1}\left( \frac{1}{2}\widetilde{R}^{\rho_k\sigma_k}_{\mu_k\nu_k} - t^2K^{\rho_k}_{\mu_k}K^{\sigma_k}_{\nu_k} \right).
\label{boundterm} 
\end{equation}
Here $ \widetilde{R}^{\rho\sigma}_{\mu\nu} $ is the intrinsic Riemann tensor of the $ \rho=\epsilon $ surface (of the metric $ H_{\mu\nu} $) and $ K_{\mu\nu} $ is its extrinsic curvature surface along the normal direction $ \rho $. The Chern form \eqref{boundterm} is the Gibbons--Hawking term of pure Lovelock gravity.

Taking the $ \epsilon\rightarrow 0 $ limit, we see that the boundary terms lead to a localized $ \alpha $-dependent contribution at $ A $ which should match with $ D_{(m,p)}(\alpha) $ once integrated over $ \alpha $. To compute this contribution, it is useful to scale the radial coordinate $ \rho \rightarrow \alpha \rho $ so that the near defect metric becomes
\begin{equation}
ds^2 = \rho^2 d\Omega_{2p-1}^2 + \frac{1}{\alpha^2}d\rho^2 + h_{ij}(x)dx^idx^j.
\label{nearcoords}
\end{equation}
In these coordinates $ \partial_\alpha H_{\mu \nu} = 0 $ so that the boundary term is a total derivative. Integrating from $ \alpha=1 $, we get
\begin{equation}
D_{(m,p)}(\alpha) = -\lim_{\epsilon\rightarrow 0}\int_{\partial \mathcal{D}_\epsilon}\sqrt{H}\,\bigl( B_{(m)} - B_{(m)}\lvert_{\alpha=1} \bigr).
\label{Dchern}
\end{equation}
This limit is computed in Appendix \ref{app:chern} and the result matches with the formula \eqref{master} obtained using the regularization method. Note that the expression \eqref{Dchern} holds only in the coordinates \eqref{nearcoords}.

One often regularizes manifolds containing singularities by cutting holes around them and, in that case, one has to introduce boundary terms at the holes. Heuristically, the fact that the defect contribution also arises from boundary terms \eqref{Dchern} ensures that cutting holes is equivalent to smoothing out the singularities.

We also note an interesting similarity of the formula \eqref{Dchern} with the ADM mass. For $ m=1 $, the boundary Chern form $ B_{(1)} \propto K $ is the trace of the extrinsic curvature of $ \partial \mathcal{D}_\epsilon $. In that case, the formula \eqref{Dchern} is similar to a formula for the ADM mass in Einstein gravity \cite{hawking_gravitational_1996}
\begin{equation}
M_{\text{ADM}} = \lim_{r\rightarrow \infty}\int \sqrt{\sigma}\, (K - K\lvert_{(0)})
\end{equation}
where the integral is over a large sphere of radius $ r $ at spatial infinity and $ K $ is the trace of the extrinsic curvature of the sphere. $ K\lvert_{(0)} $ is computed in a background spacetime that does not contain the massive object. It is possible that the formulas are related in the context of brane solutions where the defect formula could be used to compute mass.

\section{Codimension-even defects in Lovelock--Chern--Simons gravity}\label{sec:LCS}

For torsionless (Riemannian) geometries $ \mathcal{M} $, Lovelock--Chern--Simons (LCS) gravity in $ D=2m+1 $ dimensions has the action
\begin{equation}
I_{(m)}[\mathcal{M}] = \int_{\mathcal{M}}\sqrt{G}\,\mathcal{L}_{(m)}
\label{LCSaction}
\end{equation}
where
\begin{equation}
\mathcal{L}_{(m)} = \frac{\kappa}{2^{m}}\int_0^1dt\,\delta^{a_1b_1\ldots a_{m}b_{m}}_{c_1d_1\ldots c_{m}d_{m}}\prod_{n=1}^m\left(R^{c_nd_n}_{a_nb_n} + \frac{t^2}{\ell^2}\delta^{c_nd_n}_{a_nb_n} \right).
\label{lovelocklagr}
\end{equation}
The action is a Chern--Simons form for the AdS isometry group and it is traditionally written using differential forms in the first order formalism \cite{chamseddine_topological_1990} (see also \cite{cvetkovic_holography_2017}). In that case, the geometry $ \mathcal{M} $ can have non-vanishing torsion and the independent variables to be varied are the vielbein and the spin connection. In this work, we focus on the torsionless sector of the theory with a variational principle for the metric.

By expanding the product as a binomial series, performing the Kronecker contractions and integrating term by term over $ t $ one obtains
\begin{equation}
I_{(m)}[\mathcal{M}] = \kappa\int_{\mathcal{M}}\sqrt{G}\,\sum_{n=0}^m c_{(m,n)}\mathcal{R}_{(n)}
\end{equation}
where (see also \cite{crisostomo_black_2000})\,\footnote{The Kronecker contractions give a factor of $ (2m-2n+1)! $ while the corresponding $ t $-integral gives a factor of $ (2m-2n+1)^{-1} $.}
\begin{equation}
c_{(m,n)} = (2m-2n)!\,\binom{m}{n}\left( \frac{1}{\ell^2}\right)^{m-n}.
\label{cs}
\end{equation}
Lovelock--Chern--Simons gravity is thus a special case of Lovelock gravities. The theory has a unique AdS vacuum which can be seen from the equations of motion. The variation of a single Lovelock scalar $ \mathcal{R}_{(n)} $ gives 
\begin{equation}
E^a_{b(n)} =- \frac{1}{2}\frac{1}{2^{n}}\delta^{aa_1b_1\ldots a_nb_n}_{bc_1d_1\ldots c_nd_n}R^{c_1d_1}_{a_1b_1}\cdots R^{c_nd_n}_{a_nb_n}
\label{singleeom}
\end{equation}
so that the equations of motion tensor of the action \eqref{lovelocklagr} is
\begin{equation}
\sum_{n=0}^{m}c_{(m,n)}E^a_{b(n)}=-\frac{1}{2}\frac{1}{2^{m}}\delta^{aa_1b_1\ldots a_mb_m}_{bc_1d_1\ldots c_md_m}\prod_{n=1}^m\left( R^{c_nd_n}_{a_nb_n} + \frac{1}{\ell^2} \delta^{c_nd_n}_{a_nb_n}\right).
\label{luveoms}
\end{equation}
This equality is non-trivial and is a result of the particular form of the parameters $ c_{(m,n)} $.\footnote{The equality can be proven by expanding right hand side as a binomial series which gives the factor of $ \binom{m}{n} $. The remaining Kronecker contractions give the factor of $ (2m-2n)! $ matching with the left hand side.} From \eqref{luveoms} is now clear that there is a unique AdS vacuum with curvature $ -1\slash \ell^2 $. Hence LCS gravity is an example of a Lovelock Unique Vacuum theory \cite{crisostomo_black_2000,kastor_black_2006}.

We introduce the AdS curvature tensor \cite{mora_transgression_2006}
\begin{equation}
\mathcal{F}^{cd}_{ab}  = R^{cd}_{ab}  + \frac{1}{\ell^{2}} \delta^{cd}_{ab}
\label{starredriem}
\end{equation}
that measures curvature deviations from pure AdS space. Then the equations of motion \eqref{luveoms} can be written as
\begin{equation}
\mathcal{E}_{b(m)}^a = 0
\label{lcseom}
\end{equation}
where $ \mathcal{E}_{b(m)}^a $ is the tensor \eqref{singleeom} with all Riemann tensors replaced by the AdS curvature \eqref{starredriem}.

Since the LCS action is a Chern--Simons action, the solutions of the theory correspond to flat connections of the AdS isometry group in the first order formalism. In the metric formalism, the topological nature of the theory is manifested in a local condition that all the solutions satisfy. The condition follows from the relation
\begin{equation}
(2m)!^{2}\,\delta_{aa_1b_1\ldots a_mb_m}^{bc_1d_1\ldots c_md_m}\mathcal{E}_{b(m)}^a = \mathcal{F}^{[c_1d_1}_{[a_1b_1} \cdots\, \mathcal{F}^{c_md_m]}_{a_mb_m]} \equiv \mathcal{F}^{c_1d_1\ldots c_md_m}_{a_1b_1\ldots a_mb_m(m)}
\end{equation}
which follows from the identity
\begin{equation}
\delta^{bb_1 \ldots b_{2m}}_{aa_1\ldots a_{2m}}\delta^{ac_1\ldots c_{2m}}_{bd_1\ldots d_{2m}} = \delta^{c_1\ldots c_{2m}}_{a_1\ldots a_{2m}}\delta^{b_1\ldots b_{2m}}_{d_1\ldots d_{2m}}
\end{equation}
valid only in $ D = 2m+1 $.\footnote{The identity can be proven using $ \delta^{bb_1 \ldots b_{2m}}_{aa_1\ldots a_{2m}} = \epsilon^{bb_1 \ldots b_{2m}}\epsilon_{aa_1\ldots a_{2m}} $ and $ \epsilon^{bb_1 \ldots b_{2m}}\epsilon_{bd_1\ldots d_{2m}} = \delta^{b_1\ldots b_{2m}}_{d_1\ldots d_{2m}} $ that only hold in $ D=2m+1 $.} The equations of motion then imply that all the solutions satisfy
\begin{equation}
\mathcal{F}^{c_1d_1\ldots c_md_m}_{a_1b_1\ldots a_mb_m(m)} = 0.
\label{lovelockads}
\end{equation}
We call such solutions locally Lovelock--AdS. When the AdS curvature $ \ell \rightarrow \infty $, Lovelock--Chern--Simons gravity reduces to pure Lovelock gravity in $ D = 2m+1 $. From \eqref{lovelockads} it follows that the solutions of that theory are Lovelock flat \cite{dadhich_lovelock_2012,kastor_riemann-lovelock_2012}.

\subsection{Holographic Weyl anomaly of Lovelock--Chern--Simons gravity}

Consider Euclidean AdS$ _{2m+1} $ with Lovelock--Chern--Simons gravity in the bulk. In \cite{banados_holographic_2006,cvetkovic_holography_2017} it was shown that the Fefferman--Graham expansion of solutions of LCS gravity is finite. In other words, the asymptotic behaviour of solutions takes the form\,\footnote{The patch covered by the coordinates \eqref{FGgauge} does not necessarily extend beyond the asymptotic region to cover the whole manifold \cite{banados_holographic_2006}.}
\begin{equation}
ds^{2} = \frac{\ell^{2}}{z^{2}}(dz^{2} + \gamma_{\mu\nu}dx^{\mu}dx^{\nu})
\label{FGgauge}
\end{equation}
where
\begin{equation}
\gamma_{\mu\nu}(x,z) = g_{\mu\nu(0)}(x) + z^{2}g_{\mu\nu(1)}(x) + z^{4}g_{\mu\nu(2)}(x)
\end{equation}
and $ \mu,\nu,\ldots $ run over the $ 2m $ boundary coordinates. Here $ g_{(0)} $ is the metric on the conformal boundary $ \mathcal{B} $ and it is defined up to a Weyl transformation.
 
Let $ I_{(m)}^{\text{reg}}[\mathcal{M}] $ be the regularized on-shell LCS action of a solution $ \mathcal{M} $ of the equations of motion (which are all locally Lovelock--AdS). It is defined by restricting the integral in the on-shell action to end on the cut-off surface $ z=\epsilon $. The renormalized on-shell action $ I_{(m)}^{\text{ren}}[\mathcal{M}] $ is
\begin{equation}
I_{(m)}^{\text{ren}}[\mathcal{M}] = \lim_{\epsilon\rightarrow 0}\,\bigl(I_{(m)}^{\text{reg}}[\mathcal{M}] + I_{(m)}^{\text{ct}}\bigr)
\end{equation}
where $ I_{(m)}^{\text{ct}} $ are counterterms that are integrals on the cut-off surface $ z=\epsilon $. The boundary stress-energy tensor is then defined as (see for example \cite{de_haro_holographic_2000})
\begin{equation}
\tau_{\mu\nu(m)} = \frac{2}{\sqrt{g_{(0)}}}\frac{\delta I_{(m)}^{\text{ren}}[\mathcal{M}]}{\delta g_{(0)}^{\mu\nu}} = \lim_{\epsilon \rightarrow 0}\frac{2}{\sqrt{\gamma(x,\epsilon)}}\frac{\delta I_{(m)}^{\text{ren}}[\mathcal{M}]}{\delta \gamma^{\mu\nu}(x,\epsilon)}
\end{equation}
and it is a function of the conformal representative $ g_{(0)} $ only.

The holographic Weyl anomaly is the non-vanishing of the trace of $ \tau_{\mu\nu(m)} $ which measures the response of the on-shell action with respect to Weyl transformations $ g_{(0)} \rightarrow \Omega^{2}g_{(0)} $ of the boundary metric. The tensor $ \tau_{\mu\nu(m)} $ was computed in the first order formulation of LCS gravity in \cite{banados_chern-simons_2004,banados_counterterms_2005,banados_holographic_2006,cvetkovic_holography_2017} and the resulting holographic Weyl anomaly is given by
\begin{equation}
\tau_{\,\mu(m)}^{\mu} = \kappa\ell\,\mathcal{R}_{(m)}[g_{(0)}]
\label{weyl}
\end{equation}
where $ \mathcal{R}_{(m)}[g_{(0)}] $ is the Lovelock scalar of the boundary metric. This translates to an expansion of the regularized on-shell action:
\begin{equation}
I^{\text{reg}}_{(m)}[\mathcal{M}] = \int_{\mathcal{M}, z\geq \epsilon} \sqrt{G}\,\mathcal{L}_{(m)} = \int_{\mathcal{B}} \sqrt{g_{(0)}}\,\kappa\ell\,\mathcal{R}_{(m)}[g_{(0)}]\,\log{\frac{L}{\epsilon}} + \ldots
\label{actexp}
\end{equation}
where $ G $ is the metric of $ \mathcal{M} $, $ L $ is a length scale associated with the boundary metric $ g_{(0)} $ and dots contain non-universal power law divergences (that are subtracted in the renormalized action).

Assuming a holographic duality involving LCS gravity existed, the renormalized on-shell action would be related to a partition function $ Z[\mathcal{B}] $ of a non-unitary CFT on the boundary as
\begin{equation}
I_{(m)}^{\text{ren}}[\mathcal{M}] = -\log{Z[\mathcal{B}]}
\label{duality}
\end{equation}
in the saddle-point approximation. Then $ \tau_{\mu\nu(m)} $ would compute the expectation value of the CFT stress-tensor $ \langle T_{\mu\nu}\rangle_{g_{(0)}} $ on the background $ g_{(0)} $ and \eqref{weyl} translates to the Weyl anomaly of the boundary CFT. Whether a holographic duality involving LCS gravity exists is not relevant to us, because all our computations are classical and independent of a quantized duality.

\subsection{Partition functions of codimension-even defects}

Our goal is to study the Weyl anomaly \eqref{weyl} in the presence of a codimension-$ 2p $ defect on the boundary. Because it is given in terms of the Lovelock scalar, we will be able to use the defect formula \eqref{master} to compute the contribution coming from the defect. This is done without any reference to the gravity action and is the same as computing the partition function of a non-unitary CFT with purely type-A Weyl anomaly.

So let us consider a $ 2m $-dimensional CFT with the anomaly \eqref{weyl} and place it on a background $ \mathcal{B} $. Let $ A_{2m-2p} $ be a codimension-$ 2p $ surface embedded in $ \mathcal{B} $ and assume that the surface is characterized by a single length scale $ R $. A simple example is a sphere $ A_{2m-2p} = S_R^{2m-2p} $ of radius $ R $ embedded in $ \mathcal{B} = \mathbb{R}^{2m} $ which will be our focus in section \ref{sec:sphere}.

We introduce a deficit solid angle parametrized by $ \alpha $ along the surface $ A_{2m-2p} $ which defines a boundary geometry $ \mathcal{B}_{\alpha} $ containing a codimension-$ 2p $ defect $ A_{2m-2p} $. Using the Weyl anomaly \eqref{weyl}, we can compute the response of the partition function to a scale transformation:\,\footnote{See \cite{myers_holographic_2011,hung_holographic_2011} for a similar approach to computing entanglement entropy using the Weyl anomaly.}
\begin{equation}
R\frac{d}{dR} \log{Z[\mathcal{B}_\alpha]} = -\int_{\mathcal{B}_{\alpha}}d^{2m}x \sqrt{g}\,\langle T^{\mu}_{\;\;\mu}\rangle = -\int_{\mathcal{B}_{\alpha}}d^{2m}x \sqrt{g}\,\kappa\ell\,\mathcal{R}_{(m)}[g].
\label{cftweyl}
\end{equation}
where $ g $ is the metric of $ \mathcal{B}_{\alpha} $. Using \eqref{master} for the contribution of the defect, we get
\begin{equation}
R\frac{d}{dR} \log{Z[\mathcal{B}_\alpha]} =- C_{(m,p)}U_{(p)}(\alpha)\int_{A_{2m-2p}}\sqrt{\sigma}\, \kappa\ell\,\widehat{\mathcal{R}}_{(m-p)}^{\text{defect}} + R\frac{d}{dR} \log{Z[\mathcal{B}_\alpha\backslash A]}
\label{ap}
\end{equation}
where $ \sigma $ is the induced metric of $ A_{2m-2p} $ and $ \widehat{\mathcal{R}}_{(m-p)}^{\text{defect}} $ is its Lovelock scalar. There is an extra contribution coming from the region outside of the defect, because $ \mathcal{R}_{(m)}[g] $ does not necessarily vanish there. Since the integral of the Lovelock scalar is scale invariant, it produces a logarithmic divergence when integrated over $ R $:
\begin{equation}
\log{Z[\mathcal{B}_\alpha]} = -C_{(m,p)}U_{(p)}(\alpha)\int_{A_{2m-2p}}\sqrt{\sigma}\, \kappa\ell\,\widehat{\mathcal{R}}_{(m-p)}^{\text{defect}}\,\log{\frac{R}{\epsilon}}+\log{Z[\mathcal{B}_\alpha\backslash A]}
\label{defectZ}
\end{equation}
where $ \epsilon $ is the UV cut-off the CFT. This result can be equivalently stated in terms of the renormalized bulk LCS action \eqref{duality}. Next we will show how the logarithmic piece is obtained starting from the on-shell brane action in the bulk.

\subsection{On-shell actions of codimension-even branes}

In the previous section, we computed the partition function of a defect on the conformal boundary by using the anomaly \eqref{weyl}. Given the dual geometry $ \mathcal{M}_{\alpha} $, this translates to a logarithmic divergence in $ I^{\text{ren}}_{(m)}[\mathcal{M}_{\alpha}] $ which should be directly computable starting from the action itself. A dual geometry $ \mathcal{M}_{\alpha} $ that asymptotes to the defect geometry $ \mathcal{B}_{\alpha} $ on the boundary contains a codimension-$ 2p $ brane $ \Sigma_{2m-2p+1} $ anchored to the defect $ A_{2m-2p} $ ($ \partial \Sigma = A $). By a brane we mean a surface with a solid angle deficit parametrized by $ \alpha $ at each point. Hence we can use the defect formula \eqref{master} to compute the corresponding brane contribution to the action and we find that it indeed reproduces the defect contribution of \eqref{defectZ}.

The LCS action is a linear combination $ \sum_nc_{(m,n)} \mathcal{R}_{(n)} $ up to $ n=m $. By the defect formula \eqref{master}, only Lovelock scalars with $ n \geq m-p $ contribute to the localized contribution of the brane. The resulting regularized on-shell action becomes
\begin{equation}
I_{(m)}^{\text{reg}}[\mathcal{M}_\alpha] =  U_{(p)}(\alpha)\sum_{n=p}^{m} c_{(m,n)}  C_{(n,p)}\int_{\Sigma_{2m-2p+1}}\sqrt{h}\,\kappa\,\widehat{\mathcal{R}}_{(n-p)} + I_{(m)}^{\text{reg}}[\mathcal{M}_\alpha \backslash\, \Sigma]
\label{firstsub}
\end{equation}
where $ h $ is the induced metric of the brane and $ I_{(m)}^{\text{reg}}[\mathcal{M}_\alpha \backslash\, \Sigma] $ is the action computed with the regular part of the solution. Changing the summation variable as $ n\rightarrow n-p $, the sum in \eqref{firstsub} becomes
\begin{equation}
\sum_{n=0}^{m-p} c_{(m,n+p)}  C_{(n+p,p)}\int_{\Sigma}\sqrt{h}\,\widehat{\mathcal{R}}_{(n)}.
\label{sum}
\end{equation}
The parameters $ c_{(m,n)} $ \eqref{cs} and $ C_{(m,p)} $ \eqref{combcoeff} satisfy
\begin{equation}
c_{(m,n+p)} = c_{(m-p,n)}\frac{m!n!}{(m-p)!(n+p)!}, \quad C_{(n+p,p)} = C_{(m,p)}\frac{(m-p)!(n+p)!}{m!n!}
\end{equation}
so that they obey the remarkable identity
\begin{equation}
c_{(m,n+p)} C_{(n+p,p)} = c_{(m-p,n)} C_{(m,p)}.
\end{equation}
Thus the coefficient $ C_{(m,p)} $ can be moved out of the sum \eqref{sum} and the on-shell action becomes
\begin{equation}
I_{(m)}^{\text{reg}}[\mathcal{M}_\alpha] = C_{(m,p)}U_{(p)}(\alpha)\int_{\Sigma}\sqrt{h}\,\widehat{\mathcal{L}}_{(m-p)} + I_{(m)}^{\text{reg}}[\mathcal{M}_\alpha \backslash\, \Sigma]
\label{on-shell}
\end{equation}
where $ \widehat{\mathcal{L}}_{(m-p)} $ is the intrinsic LCS Lagrangian of the brane:
\begin{equation}
\widehat{\mathcal{L}}_{(m-p)}=\kappa\sum_{n=0}^{m-p}c_{(m-p,n)}  \widehat{\mathcal{R}}_{(n)} = \frac{\kappa}{2^{m-p}}\int_0^1dt\,\delta^{i_1j_1\ldots i_{m-p}j_{m-p}}_{k_1l_1\ldots k_{m-p}l_{m-p}}\prod_{n=1}^{m-p}\left(\widehat{R}^{k_nl_n}_{i_nj_n} + \frac{t^2}{\ell^2}\delta^{k_nl_n}_{i_nj_n} \right).
\label{intrlagr}
\end{equation}
The latin indices run over the $ 2m-2p+1 $ brane coordinates. The result can be equivalently written as
\begin{equation}
\int_{\mathcal{M}_{\alpha}}\sqrt{G}\,\mathcal{L}_{(m)} = C_{(m,p)}U_{(p)}(\alpha)\int_{\Sigma}\sqrt{h}\,\widehat{\mathcal{L}}_{(m-p)} +\int_{\mathcal{M}_\alpha \backslash\, \Sigma}\sqrt{G}\,\mathcal{L}_{(m)}
\end{equation}
and the brane contribution is simply LCS action localized on the brane.

In the above derivation, we assumed that we had found a solution $ \mathcal{M}_{\alpha} $ that contains a surface of conical singularities $ \Sigma $ with an induced metric $ h $. To generate such conical solutions in the first place, we introduce an action which is a sum of the LCS action and an auxiliary brane action:\,\footnote{Same idea is used in \cite{dong_gravity_2016}.}
\begin{equation}
-C_{(m,p)}U_{(p)}(\alpha)\int_{\Sigma}\sqrt{h}\,\widehat{\mathcal{L}}_{(m-p)} +\int_{\mathcal{M}}\sqrt{G}\,\mathcal{L}_{(m)}
\label{auxaction}
\end{equation}
where the integral over $ \mathcal{M} $ includes $ \Sigma $. The metric $ G $ of $ \mathcal{M} $ and the embedding functions of the brane $ \Sigma $ (location of the brane) constitute the set of parameters to be varied and solved from the equations of motion. The equation for $ G $ contains a delta function source which leads to conical singularities of strength $ \alpha $ along $ \Sigma $ in the solution. This can be seen at the level of the action: adding singularities along $ \Sigma $ in $ \mathcal{M} $ produces an extra term that cancels the auxiliary brane action. Since the action \eqref{auxaction} depends on the embedding functions only through the induced metric $ h $, the resulting equation of motion for the functions is
\begin{equation}
\widehat{\mathcal{E}}^{i}_{j(m-p)}[h] = 0.
\label{inducedLCS}
\end{equation}
In other words, the induced metric $ h $ of the conical surface is a solution of lower dimensional LCS gravity \eqref{lcseom}.\footnote{Proving this at the level of equations of motion might require an analysis similar to \cite{boisseau_dynamics_1997}.}

We can now use \eqref{inducedLCS} to expand the first term in the on-shell action \eqref{on-shell}. Because $ \Sigma $ is anchored to the conformal boundary, $ h $ is an asymptotically locally AdS solution of LCS gravity so that it has a truncated Fefferman--Graham expansion similarly to $ G $. Hence the contribution from the conical surface has the same expansion \eqref{actexp} as the full action:
\begin{equation}
\int_{\Sigma}\sqrt{h}\,\widehat{\mathcal{L}}_{(m-p)} = \int_A \sqrt{\sigma}\,\kappa\ell\, \widehat{\mathcal{R}}^{\text{defect}}_{(m-p)}\,\log{\frac{R}{\epsilon}} + \ldots
\end{equation}
where $ A = \partial \Sigma $ is the boundary defect, $ R $ is the length scale associated with $ A $ and the dots denote non-universal power law divergences. After renormalization of both the brane and the full action, the on-shell action \eqref{on-shell} becomes
\begin{equation}
I_{(m)}^{\text{ren}}[\mathcal{M}_\alpha] =C_{(m,p)}U_{(p)}(\alpha)\int_A \sqrt{\sigma}\,\kappa\ell\, \widehat{\mathcal{R}}^{\text{defect}}_{(m-p)}\,\log{\frac{R}{\epsilon}} + I_{(m)}^{\text{ren}}[\mathcal{M}_\alpha \backslash\, \Sigma]
\end{equation}
and using $ I_{(m)}^{\text{ren}}[\mathcal{M}_\alpha] = -\log{Z[\mathcal{B}_{\alpha}]} $ the defect contribution matches with the CFT computation \eqref{defectZ}. The part of the action not coming from the defect reproduces $ \log{Z[\mathcal{B}_\alpha\backslash A]} $ which follows from the holographic Weyl anomaly \eqref{weyl} of the full action.

It is remarkable that the on-shell brane action is Lovelock--Chern--Simons gravity of lower dimension localized on the brane. The exact localization is expected, because the Weyl anomaly localizes on defects on the boundary. Since $ \mathcal{L}_{(m)} $ gives the holographic anomaly $ \mathcal{R}_{(m)} $, the only brane Lagrangian that produces $ \widehat{\mathcal{R}}^{\text{defect}}_{(m-p)} $ holographically is $ \widehat{\mathcal{L}}_{(m-p)} $. The matching of the two computations thus provides a strong consistency check of the defect formula \eqref{master}.

\section{Duality between spherical defects and hyperbolic branes}\label{sec:sphere}

In this section, we explicitly demonstrate the duality between codimension-$ 2p $ defects and branes proven in previous sections for the case of a spherical defect on the boundary. We show that the dual solution of the defect is a brane with hyperbolic intrinsic geometry and find that the logarithmic divergence in the on-shell action matches with the partition function.

\subsection{Partition function of a spherical defect}

\begin{figure}[t]
	\begin{subfigure}[t]{0.5\textwidth}
		\centering
		\includegraphics{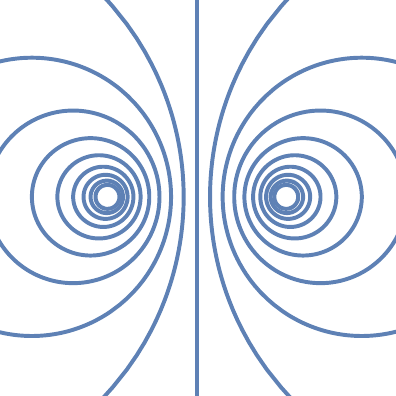}
		\subcaption{}
	\end{subfigure}
	\begin{subfigure}[t]{0.5\textwidth}
		\centering
		\includegraphics{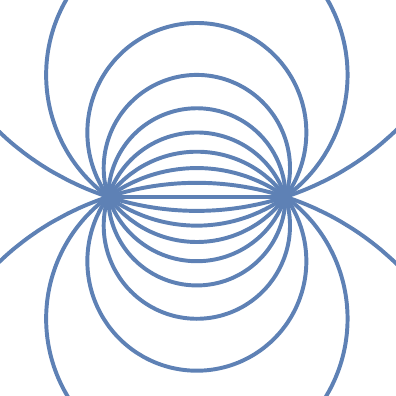}
		\subcaption{}
	\end{subfigure}
	\caption{Visualization of the coordinates $ u $ and $\theta $ \eqref{confmap} for $ m=p=1 $ and metric $ ds^{2} = d\rho^{2} + d\tilde{\rho}^{2} $: $ (a) $ constant-$ u $ slices which are circles $ (b) $ constant-$ \theta $ slices. In this case, the defect will be an $ S^{0} $ which consists of two conical singularities at $ \tilde{\rho} = \pm R $.}
	\label{bipolar}
\end{figure}

Consider a non-unitary CFT with the Weyl anomaly \eqref{weyl} on $ \mathbb{R}^{2m} $. In this section, we will compute the partition function of a spherical defect $ S^{2m-2p}_{R} \subset \mathbb{R}^{2m} $ of radius $ R $ using the Weyl anomaly. To construct the metric of the defect, we start from $ \mathbb{R}^{2m} = \mathbb{R}^{2p-1}\times \mathbb{R}^{2m-2p+1} $ and write each factor in spherical coordinates:
\begin{equation}
ds^{2}_{\mathbb{R}^{2m}} = d\rho^{2} + \rho^{2}d\Omega^{2}_{2p-2}+  d\tilde{\rho}^{2} + \tilde{\rho}^{2}d\tilde{\Omega}^{2}_{2m-2p}
\label{flat}
\end{equation}
with coordinate ranges such that all of $ \mathbb{R}^{2m} $ is covered. We parametrize the sphere as the surface
\begin{equation}
\tilde{\rho}^{2} = R^{2}, \quad \rho=0
\label{sphere}
\end{equation}
embedded inside the factor $ \mathbb{R}^{2m-2p+1} $. Perform now the transformation
\begin{equation}
\rho = \frac{R\,\sin{\theta}}{\cosh{u} + \cos{\theta}}, \quad \tilde{\rho} = \frac{R\,\sinh{u}}{\cosh{u} + \cos{\theta}}
\label{confmap}
\end{equation}
with rest of the coordinates kept fixed. The metric \eqref{flat} becomes
\begin{equation}
ds^{2}_{\mathbb{R}^{2m}} = \frac{R^2}{(\cosh{u} + \cos{\theta})^{2}}\left( d\theta^{2} + \sin^{2}{\theta}\,d\Omega^2_{2p-2} + du^{2} + \sinh^{2}{u}\, d\tilde{\Omega}_{2m-2p}^2\right).
\label{tori}
\end{equation}
with the metric in brackets being $ S^{2p-1}\times \mathbb{H}^{2m-2p+1} $. The ranges of the coordinates are
\begin{align}
0&\leq \theta < \pi, & 0&\leq u, &  m&>p\nonumber\\
0&\leq \theta < \pi, & -\infty &< u <\infty, &  m&=p>1\nonumber\\
0&\leq \theta < 2\pi, & -\infty &< u <\infty, &  m&=p=1.
\end{align}
See figure \ref{bipolar} for a visualization of the coordinates for $ m=p=1 $. The case $ m=p>1 $ is similar, but instead of an $ S^{1} $ shrinking to zero size at $ \tilde{\rho} = \pm R $ it is an $ S^{2m-1} $ that shrinks.

By inverting the conformal factor, \eqref{tori} can also be written as
\begin{equation}
d\Omega^2_{2p-1} + d\Sigma_{2m-2p+1}^2 = \frac{4R^{2}}{[\rho^{2} + (\tilde{\rho}-R)^{2}][\rho^{2} + (\tilde{\rho}+R)^{2}]}\,ds^{2}_{\mathbb{R}^{2m}}
\label{conformalflat}
\end{equation}
where $  d\Sigma_{2m-2p+1}^2 $ denotes the metric of unit $ \mathbb{H}^{2m-2p+1} $. This shows that the space $ S^{2p-1}\times \mathbb{H}^{2m-2p+1} $ is locally conformally flat and that the transformation \eqref{confmap} is a conformal map from $ \mathbb{R}^{2m} $ to $ S^{2p-1}\times \mathbb{H}^{2m-2p+1} $.\footnote{The conformal flatness of $ S^{2p-1}\times \mathbb{H}^{2m-2p+1} $ is a special case of a more general theorem: a non-flat Riemannian manifold, which is locally a direct product space, is locally conformally flat if and only if it is locally equal to $ \Sigma(R)\times [a,b] $ or $ \Sigma(R) \times \Sigma(-R) $ \cite{brozos-vazquez_complete_2006}. Here $ [a,b]\subset \mathbb{R} $ is an interval and $ \Sigma(R) $ is a space of constant curvature $ R $.} Therefore it is a generalization of the Euclidean Casini--Huerta--Myers map \cite{casini_towards_2011} which is a conformal map from $ \mathbb{R}^{2m} $ to $ S^{1}\times \mathbb{H}^{2m-1} $.

From \eqref{conformalflat} we see that the conformal factor diverges along the sphere \eqref{sphere} so that it is mapped to $ u=\infty $ in the new coordinates (the interior of the sphere is mapped to $ \mathbb{H}^{2m-2p+1} $). Hence we can introduce a deficit solid angle $ \alpha $ along the sphere as
\begin{equation}
ds^{2} = \frac{R^2}{(\cosh{u} + \cos{\theta})^{2}}\left( \alpha^{2}d\Omega^2_{2p-1} + d\Sigma_{2m-2p+1}^2\right).
\label{sphericaldefect}
\end{equation}
As we approach the sphere $ u \rightarrow \infty $, the metric \eqref{sphericaldefect} behaves as
\begin{equation}
ds^{2}  = \tilde{u}^{2}\alpha^{2}d\Omega^2_{2p-1} + R^{2} d\tilde{\Omega}_{2m-2p}^{2}
\end{equation}
where $ \tilde{u} = Re^{-u} $ and from which we see that there is a conical singularity at $ \tilde{u} = 0 $. Hence \eqref{sphericaldefect} is the metric of a spherical defect of radius $ R $ and we denote it by $ \mathcal{B}_{\alpha} $.

We can now use the formula \eqref{defectZ} to compute the partition function of the defect. The sphere has constant curvature tensor $ \widehat{R}^{kl}_{ij} = (1\slash R^2)\delta^{kl}_{ij} $ so that performing the Kronecker delta contractions yields
\begin{equation}
\widehat{\mathcal{R}}_{(m-p)}^{\text{defect}} = \frac{(2m-2p)!}{R^{2m-2p}}.
\end{equation}
The integral over the sphere produces a factor of $ R^{2m-2p} $ exactly cancelling the corresponding one in the denominator and we get
\begin{equation}
\int_{S_R^{2m-2p}}\sqrt{\sigma}\, \widehat{\mathcal{R}}_{(m-p)}^{\text{defect}} = \Omega_{2m-2p}(2m-2p)!.
\end{equation}
The cancellation of $ R $ is expected due to scale invariance of this expression. The CFT partition function \eqref{defectZ} becomes
\begin{equation}
\log{Z[\mathcal{B}_\alpha]} = -\kappa\ell\,\Omega_{2m-2p}(2m-2p)!\,C_{(m,p)}U_{(p)}(\alpha)\log{\frac{R}{\epsilon}} + \log{Z[\mathcal{B}_\alpha\backslash \,S^{2m-2p}_{R}]}.
\label{CFTresult}
\end{equation}
One could explicitly compute the contribution from $ \mathcal{B}_\alpha\backslash \,S^{2m-2p}_{R} $ for the metric \eqref{sphericaldefect}, but we will not do that here. We will now show how the first term arises holographically from the action of a hyperbolic brane in the bulk.

\subsection{Euclidean hyperbolic brane solution}\label{hypbrane}

We look for codimension-$ 2p $ hyperbolic brane solutions of Lovelock--Chern--Simons gravity that are dual to the spherical defect \eqref{sphericaldefect}. Motivated by the $ S^{2p-1}\times \mathbb{H}^{2m-2p+1} $ structure on the boundary, we attempt the ansatz
\begin{equation}
ds^2 = f(r)\,\ell^{2}\alpha^{2} d\Omega^2_{2p-1} + f(r)^{-1}dr^2 + r^2d\Sigma_{2m-2p+1}^2
\label{hypansatz}
\end{equation}
where $ f(r) $ is an unknown function and $ \alpha $ is a parameter that will determine the solid angle deficit.

The equations of motion \eqref{lcseom} of LCS gravity are
\begin{equation}
\delta^{aa_1b_1\ldots a_mb_m}_{bc_1d_1\ldots c_md_m}\mathcal{F}^{c_1d_1}_{a_1b_1} \cdots \mathcal{F}^{c_md_m}_{a_mb_m} = 0.
\label{eomsum}
\end{equation}
Turns out that to determine $ f(r) $ all we need is the $ (r,r) $-component. It is given by
\begin{equation}
\delta^{r \phi_1\phi_2\ldots \phi_{2p-1}i_1\ldots i_{2m-2p}}_{r\varphi_1\varphi_2\ldots \varphi_{2p-1}j_1\ldots j_{2m-2p}}\mathcal{F}^{\varphi_1j_1}_{\phi_1i_1}\mathcal{F}^{\varphi_2\varphi_3}_{\phi_2\phi_3} \cdots \mathcal{F}^{\varphi_{2p-2}\varphi_{2p-1}}_{\phi_{2p-2}\phi_{2p-1}}\mathcal{F}^{j_2j_3}_{i_2i_3}\cdots \mathcal{F}^{j_{2m-2p-1}j_{2m-2p}}_{i_{2m-2p-1}i_{2m-2p}} = 0
\end{equation}
up to combinatorial prefactors which have been divided out. Here $ \varphi,\phi $ run over the coordinates of $ S^{2p-1} $ and $ i,j $ run over the coordinates of $ \mathbb{H}^{2m-2p+1} $. Note that no other terms appear in the sum \eqref{eomsum}, because all the indices have been used up in the Kronecker delta. This is the simplification that arises from the topological nature of the theory.

The angular and surface components sum up to an overall prefactor which can be divided out. Therefore the equations of motion are equivalent with
\begin{equation}
\delta^{\phi_1i_1}_{\varphi_1j_1}\mathcal{F}^{\varphi_1j_1}_{\phi_1i_1} = 0
\end{equation}
where
\begin{equation}
\mathcal{F}^{\varphi_1j_1}_{\phi_1i_1} = -\frac{f'(r)}{2r}\delta^{\varphi_1j_1}_{\phi_1i_1} + \frac{1}{\ell^2} \delta^{\varphi_1j_1}_{\phi_1i_1}.
\end{equation}
We get
\begin{equation}
-\frac{f'(r)}{2r} + \frac{1}{\ell^2} = 0 \Rightarrow f(r) = \frac{r^2-r_h^2}{\ell^2}
\label{fr}
\end{equation}
where $ r_h $ is an integration constant to be fixed below. One can now check that the metric \eqref{hypansatz} with $ f(r) $ given by \eqref{fr} is locally Lovelock--AdS (it is also asymptotically locally AdS). By an explicit computation one finds that
\begin{align}
\mathcal{F}^{\varphi j}_{\phi i}=\mathcal{F}^{rj}_{ri} = \mathcal{F}^{r\varphi}_{r\phi} &= 0\\
\mathcal{F}^{\varphi_1\varphi_2}_{\phi_1\phi_2} &\neq 0,\quad p-1 \text{ total}\\
\mathcal{F}^{kl}_{ij} &\neq 0, \quad m-p \text{ total}
\end{align}
so that there are a total of $ (p-1)+(m-p) = m-1 $ non-zero AdS curvature tensors and the anti-symmetrization over $ m $ of them vanishes $ \mathcal{F}^{c_1d_1\ldots c_md_m}_{a_1b_1\ldots a_mb_m(m)} = 0 $. Thus we have found a solution of Lovelock--Chern--Simons gravity by just solving one component of the equations of motion:
\begin{equation}
ds^2 = (r^2 - r_h^2) \,\alpha^{2} d\Omega^2_{2p-1} + \frac{\ell^2}{r^2-r_h^2}dr^2 + r^2d\Sigma_{2m-2p+1}^2.
\label{hypsolrh}
\end{equation}
Taking $ r\rightarrow \infty $, we find the asymptotic behaviour
\begin{equation}
ds^2 = r^{2}(\alpha^{2}d\Omega^2_{2p-1} +  d\Sigma_{2m-2p+1}^2)
\label{bound}
\end{equation}
which coincides with the metric of the spherical defect \eqref{sphericaldefect} up to Weyl rescaling.\footnote{The Weyl factor appearing in \eqref{sphericaldefect} can be recovered by an appropriate coordinate transformation if needed.} Hence the parameter $ \alpha $ in the metric ansatz is identified as the deficit parameter of the defect.

Expanding $ r = r_h + (1\slash 4)f'(r_h)  \rho^2 $, we find the near $ r=r_h $ behaviour
\begin{equation}
ds^2 = \rho^2 \alpha^2\frac{r_{h}^{2}}{\ell^{2}}\,d\Omega^2_{2p-1} + d\rho^2 + r_h^2 d\Sigma_{2m-2p+1}^2
\end{equation}
which has a conical singularity at $ r=r_{h} $. For $ \alpha = 1 $ there is no defect on the boundary and there should be no brane singularity in the dual solution either. This fixes the integration constant to $ r_{h} = \ell $ and we get the solution
\begin{equation}
ds^2 = (r^2 - \ell^2) \,\alpha^{2} d\Omega^2_{2p-1} + \frac{\ell^2}{r^2-\ell^2}dr^2 + r^2d\Sigma_{2m-2p+1}^2.
\label{hypsol}
\end{equation}
It describes a codimension-$ 2p $ brane with deficit $ \alpha $ at $ r = \ell $ with intrinsic hyperbolic geometry. For $ \alpha = 1 $ the solution is a patch of Euclidean AdS$ _{2m+1} $ which is shown explicitly in Appendix \ref{slicing}. It corresponds to a foliation by $ S^{2p-1}\times \mathbb{H}^{2m-2p+1} $-slices and the coordinates \eqref{hypsol} are a generalization of AdS--Rindler coordinates \cite{parikh_rindler-ads/cft_2012}.\footnote{This slicing has also been used to study defect CFTs in \cite{kobayashi_towards_2019}.}

\subsection{On-shell action of the hyperbolic brane}

We will now compute the on-shell action of the hyperbolic brane solution \eqref{hypsol}. The brane at $ r = \ell $ has constant negative curvature $ \widehat{R}^{ij}_{kl} = (-1\slash \ell^{2})\delta^{ij}_{kl} $ so that the intrinsic Lagrangian is\,\footnote{$ \delta^{i_1j_1\ldots i_{m-p}j_{m-p}}_{k_1l_1\ldots k_{m-p}l_{m-p}}\delta^{k_1l_1}_{i_1j_1}\cdots \delta^{k_{m-p}l_{m-p}}_{i_{m-p}j_{m-p}} = 2^{m-p}(2m-2p+1)! $}
\begin{equation}
\widehat{\mathcal{L}}_{(m-p)} = \kappa \left( -\frac{1}{\ell^{2}}\right)^{m-p} \frac{(2m-2p+1)!(2m-2p)!!}{(2m-2p+1)!!}
\end{equation}
where we used
\begin{equation}
\int_0^1dt\,
\left(1 - t^{2} \right)^{m-p} = \frac{(2m-2p)!!}{(2m-2p+1)!!}.
\end{equation}
To compute the volume $ \text{Vol}\,\mathbb{H}^{2m-2p+1} $ of the brane, we write its induced metric as
\begin{equation}
ds^{2} = \frac{\ell^{2}}{z^{2}}\left( \frac{R^{2}}{R^{2} - z^{2}}\,dz^{2} +  (R^{2} - z^{2})\,d\Omega^{2}_{2m-2p}\right).
\end{equation}
In these coordinates, $ \mathbb{H}^{2m-2p+1} $ is covered by $ z\in[0,R] $ and the conformal boundary is located at $ z=0 $ where the brane metric matches with the metric of the defect $ S_{R}^{2m-2p} $. 

The regularized volume is
\begin{equation}
\text{Vol}\,\mathbb{H}^{2m-2p+1} = \Omega_{2m-2p}\,\ell^{2m-2p+1}\int_{\epsilon\slash R}^{1} dz\,\frac{(1-z^{2})^{m-p-1\slash 2}}{z^{2m-2p+1}}
\end{equation}
where we did a change of variables $ z \rightarrow z\slash R $. Expanding the integrand as a Taylor series, the term of order $ z^{-1} $ integrates to a logarithmic divergence:
\begin{equation}
\text{Vol}\,\mathbb{H}^{2m-2p+1} =2\ell\,(-\ell^{2})^{m-p}\,\Omega_{2m-2p}\frac{(2m-2p-1)!!}{(2m-2p)!!}\,\log{\frac{R}{\epsilon}} + \ldots
\end{equation}
and the dots contain non-universal power law divergences. We get
\begin{equation}
\int_{\Sigma}\sqrt{h}\,\widehat{\mathcal{L}}_{(m-p)} = \kappa\ell \,\Omega_{2m-2p}(2m-2p)!\log{\frac{R}{\epsilon}} + \ldots
\end{equation}
After renormalizing the power law divergences using brane counterterms, we get the renormalized on-shell action \eqref{on-shell}:
\begin{equation}
I_{(m)}^{\text{ren}}[\mathcal{M}_{\alpha}] = \kappa\ell \,\Omega_{2m-2p}(2m-2p)!\,C_{(m,p)}U_{(p)}(\alpha)\log{\frac{R}{\epsilon}} +I_{(m)}^{\text{ren}}[\mathcal{M}_\alpha \backslash\, \Sigma]
\end{equation}
where the first term matches explicitly with the result obtained from the Weyl anomaly \eqref{CFTresult} after the identification $ I_{(m)}^{\text{ren}}[\mathcal{M}_\alpha] = -\log{Z[\mathcal{B}_{\alpha}]} $.

\section{Discussion and outlook}\label{sec:disc}

In this paper, we derived the contribution of a codimension-$ 2p $ conical defect to an integral of a Lovelock scalar and applied it in holographic Lovelock--Chern--Simons gravity. The type of conical singularity we considered has a solid angle deficit parametrized by $ \alpha $. We proved that the on-shell action of a codimension-$ 2p $ brane solution, which reaches the conformal boundary, computes the logarithmic divergence in the partition function of a codimension-$ 2p $ defect and showed this explicitly in an example.

We focused on conical singularities whose metric is spherically symmetric around the singularity which translates to the defect having zero extrinsic curvature. A natural generalization of the computation is to consider squashed cones for which all the $ 2p $ extrinsic curvatures $ K^{(i)} $ are turned on. For codimension-2 defects, the extra contributions from $ K^{(i)} $ do not change the end result as they combine non-trivially to give the lower order Lovelock scalar $ \widehat{\mathcal{R}}_{(m-1)} $ of the defect. This was proven in \cite{dong_holographic_2014} for small $ 1-\alpha $, and in \cite{fursaev_distributional_2013} for arbitrary $ \alpha $, but not for all $ m $. Hence it is most likely true for arbitrary $ m, \alpha $ and we expect it to hold for codimension-$ 2p $ defects as well. However, the regularization of squashed cones is more involved making the computation of section \ref{subsec:reg} more complicated.

In the gravity context, the higher dimensional cones we considered are different from two-dimensional ones, because they are not flat (their Riemann tensor is non-zero). This is the reason why geometries including such defects do not arise as vacuum solutions of pure Einstein gravity: they would require matter stress-energy to support the additional curvature surrounding the defect. But for example in theories whose equations of motion depend on curvature only through the Lovelock curvature tensor $ \mathcal{R}^{c_1d_1\ldots c_md_m}_{a_1b_1\ldots a_mb_m(m)} $, extra matter is not needed, because the cones are Lovelock flat. Hence turning on $ \alpha $ in these theories only leads to a localized delta function source in the equations of motion. For the same reason they appear as vacuum solutions in LCS gravity where enough Riemann tensors are antisymmetrized in the equations of motion.

A remarkable fact about the LCS action is that it localizes on a codimension-$ 2p $ brane exactly which follows from the form of the coefficients $ C_{(m,p)} $ appearing in the defect formula \eqref{master}. Similar localization has been seen in Lovelock gravity in \cite{charmousis_matching_2005} where brane actions of arbitrary codimension were studied using junction conditions. In those cases, the brane action is also a Lovelock action with altered coefficients. Similar result follows from the defect formula \eqref{master} for branes with solid angle deficits. It would be interesting to understand the connection between the junction condition approach to higher codimension branes and the computations of this paper.

In the holographic computations, we did not perform renormalization of the on-shell brane action explicitly which is required to remove the non-universal power law divergences that appear in the regularized action \cite{karch_holographic_2006}. In principle, the brane counterterms can be obtained from the counterterms of the full action \cite{mora_finite_2004} by the use of the defect formula \eqref{master}. This is how counterterms to Ryu--Takayanagi formula are obtained in \cite{taylor_renormalized_2016} and a similar approach works to derive Kounterterms in Einstein gravity \cite{olea_mass_2005,olea_regularization_2007,anastasiou_einstein-ads_2018,anastasiou_renormalization_2018,anastasiou_topological_2018,anastasiou_renormalized_2019}. Since the brane action is also an LCS action, one expects that the codimension-$ 2p $ counterterms take the same form as the full counterterms.

The results of this paper probe the classical and geometric aspects of a putative holographic duality between an even-dimensional non-unitary CFT and LCS gravity. However, the existence of an actual quantized version of the duality that would arise as a limit of a string theory system is up to debate. Already the vanishing of the type-B Weyl anomaly of the dual CFT is not consistent with unitarity as shown by constraints arising from conformal collider thought experiments \cite{hofman_conformal_2008}.

\section*{Acknowledgements}

The author thanks Victor Godet, Niko Jokela, Esko Keski-Vakkuri and Miika Sarkkinen for useful comments and discussions on the draft. This work is supported by the Academy of Finland grant no 
1297472 and in part by a grant from the Osk. Huttunen Foundation.

\begin{appendix}
	
\section{Proof of vanishing of the extra terms}\label{app:proof}

In this Appendix, we show that the integrals
\begin{equation}
d_{(m,n)}(x)\propto \lim_{\rho_0\rightarrow 0}\lim_{\varepsilon\rightarrow 0}\int_0^{\rho_0 \slash \varepsilon} ds\, \varepsilon^{2(p-n)} s^{2p-1} f(s),
\label{proptoapp}
\end{equation}
that appear in \eqref{propto}, vanish for $ n<p $ in the $ \rho_0,\varepsilon\rightarrow 0 $ limits. 

We can write the integral as
\begin{equation}
d_{(m,n)}(x) \propto \lim_{\rho_0 \rightarrow 0}\lim_{\varepsilon\rightarrow 0} \varepsilon^{2(p-n)}\left[ F(\rho_0\slash \varepsilon) - F(0)\right] 
\label{genericn}
\end{equation}
where $ F(s) $ is the integral function of $ s^{2p-1}f(s) $. The initial value $ F(0) $ is $ \varepsilon $-independent and is thus taken to zero by the prefactor $ \varepsilon^{2(p-n)} $ as $ \varepsilon \rightarrow 0 $. The first term is more troublesome, but it is enough to focus on the asymptotic $ s \rightarrow \infty $ behaviour of $ F(s) $ since the $ \varepsilon \rightarrow 0 $ limit is taken first.

The function $ f(s) $ is given by
\begin{equation}
f(s) = 
\begin{cases}
\frac{\left( u(s) - \alpha^2\right)^{n}}{s^{2n}u(s)^{n-1\slash 2}}\\
\frac{\dot{u}(s)\left( u(s) - \alpha^2\right)^{n-1}}{s^{2n-1}u(s)^{n+1\slash 2}}
\end{cases}
\end{equation}
where the bottom expression has $ R^{s\phi}_{s\varphi} \sim \dot{u} $ appearing in the sum \eqref{schem} while the top expression does not. For large $ s $ it goes as
\begin{equation}
f(s) \sim 
\begin{cases}
1\slash s^{2n}\\
\dot{u}(s)\slash s^{2n-1}
\end{cases}
\end{equation}
where we used $ u(s) \sim 1 $. Given any regulator $ u(s) $ that satisfies the boundary conditions \eqref{boundconds}, there exists an $ a>0 $ such that
\begin{equation}
u(s) = 1+\mathcal{O}\left( s^{-a}\right) 
\label{require}
\end{equation}
so that $ \dot{u}(s) \sim 1\slash s^{a+1} $. Then
\begin{equation}
f(s) \sim 
\begin{cases}
1\slash s^{2n}\\
1\slash s^{2n+a}
\end{cases}
\end{equation}
Thus the integral function of $ s^{2p-1}f(s) $ goes as
\begin{equation}
F(s) \sim
\begin{cases}
s^{2(p-n)}\\
s^{2(p-n)-a}
\end{cases}
\end{equation}
Therefore when $ \varepsilon \rightarrow 0 $
\begin{equation}
\varepsilon^{2(p-n)}F(\rho_0\slash \varepsilon) \sim \begin{cases}
\rho_0^{2(p-n)}\\
\varepsilon^a\rho_0^{2(p-n)-a}
\end{cases}
\end{equation}
that both go to zero at least once the second limit $ \rho_0 \rightarrow 0 $ is taken.

\section{Defect contribution as a limit of boundary terms}\label{app:chern}

We will now prove the formula
\begin{equation}
D_{(m,p)}(\alpha) = -\lim_{\epsilon\rightarrow 0}\int_{\partial \mathcal{D}_\epsilon}\sqrt{H}\,\bigl( B_{(m)} - B_{(m)}\lvert_{\alpha=1} \bigr).
\end{equation}
derived in section \ref{sec:boundary} where $ D_{(m,p)}(\alpha) $ is given in \eqref{master} and
\begin{equation}
B_{(m)} = 2m\int_0^1dt\,\delta^{\mu\mu_1\nu_1\ldots \mu_{m-1}\nu_{m-1}}_{\nu\rho_1\sigma_1\ldots \rho_{m-1}\sigma_{m-1}}K^{\nu}_{\mu}\prod_{k=1}^{m-1}\left( \frac{1}{2}\widetilde{R}^{\rho_k\sigma_k}_{\mu_k\nu_k} - t^2K^{\rho_k}_{\mu_k}K^{\sigma_k}_{\nu_k} \right).
\label{chernapp}
\end{equation}
is the Chern form. To remove clutter, we will denote
\begin{equation}
\mathcal{Q}^{\rho\sigma}_{\mu\nu} = \widetilde{R}^{\rho\sigma}_{\mu\nu} - 2t^2K^{\rho}_{[\mu}K^{\sigma}_{\nu]}.
\end{equation}
The metric near the defect $ A $ is of the form
\begin{equation}
ds^2 = \rho^2 d\Omega_{2p-1}^2 + \frac{1}{\alpha^2}d\rho^2 + h_{ij}(x)dx^idx^j.
\end{equation}
so that the only non-zero components of the Riemann tensor and the extrinsic curvature are
\begin{equation}
\widetilde{R}^{\rho\sigma}_{\mu\nu} = \frac{1}{\rho^2}\delta^{\phi_1\phi_2}_{\varphi_1\varphi_2},\quad K_{\varphi}^\phi = \frac{\alpha}{\rho} \, \delta^\phi_{\varphi}
\end{equation}
which implies
\begin{equation}
\mathcal{Q}^{\phi_1\phi_2}_{\varphi_1\varphi_2}=\frac{1}{\rho^2}\left( 1 - \alpha^2t^2\right) \delta^{\phi_1\phi_2}_{\varphi_1\varphi_2}, \quad \mathcal{Q}^{i_1i_2}_{j_1j_2} = \widetilde{R}^{i_1i_2}_{j_1j_2} = \widehat{R}^{i_1i_2}_{j_1j_2}.
\end{equation}
where we used the Gauss-Codazzi equation and spherical symmetry to write $ \widetilde{R}^{i_1i_2}_{j_1j_2} $ in terms of the Riemann tensor $ \widehat{R}^{i_1i_2}_{j_1j_2} $ of the metric $ h $. We factorize the Kronecker sum in \eqref{chernapp} as
\begin{align}
B_{(m)} = 2m\binom{m-1}{p-1}\frac{1}{2^{m-1}}\int_0^1dt\,\delta^{\phi_1 \phi_2\ldots \phi_{2p-1}}_{\varphi_1 \varphi_2\ldots \varphi_{2p-1}}\delta^{i_{1}\ldots i_{2m-2p}}_{j_{1}\ldots j_{2m-2p}}&K^{\varphi_1}_{\phi_2} \overbrace{\mathcal{Q}_{\phi_2\phi_3}^{\varphi_2\varphi_3}\cdots \mathcal{Q}_{\phi_{2p-2}\phi_{2p-1}}^{\varphi_{2p-2}\varphi_{2p-1}}}^{p-1}\times\nonumber\\
&\times\underbrace{\mathcal{Q}^{j_{1}j_{2}}_{i_{1}i_{2}}\cdots \mathcal{Q}^{j_{2m-2p-1}j_{2m-2p}}_{i_{2m-2p-1}i_{2m-2p}}}_{m-p} + \ldots
\label{Bleading}
\end{align}
where the dots contain terms with less than $ p-1 $ angular tensors $ \mathcal{Q}^{\phi_1\phi_2}_{\varphi_1\varphi_2} $. At $ \rho = \epsilon $, we then get
\begin{equation}
B_{(m)}\lvert_{\rho = \epsilon} = \widetilde{C}_{(m,p)}\frac{1}{\epsilon^{2p-1}}\widehat{R}_{(m-p)}\int_0^1dt\,\alpha\left( 1 - \alpha^2t^2\right)^{p-1} + \mathcal{O}\left( \frac{1}{\epsilon^{2p-3}}\right)
\end{equation}
where the combinatorial factor is defined in \eqref{comb}. Performing the angular integrals by spherical symmetry gives
\begin{equation}
\int_{\partial \mathcal{D}_\epsilon}\sqrt{H} = \Omega_{2p-1}\epsilon^{2p-1}\int_{A}\sqrt{h}
\end{equation}
so that
\begin{equation}
\int_{\partial \mathcal{D}_\epsilon}\sqrt{H}\,B_{(m)} = \Omega_{2p-1}\widetilde{C}_{(m,p)}\int_0^\alpha du\,\left( 1 - u^2\right)^{p-1}\int_{A}\sqrt{h}\,\widehat{R}_{(m-p)} + \mathcal{O}(\epsilon^2)
\end{equation}
where we did the change of variables $ u = \alpha t $ in the integral. Now the difference $ B_{(m)} - B_{(m)}\lvert_{\alpha=1}  $ is proportional to the integral
\begin{equation}
\int_0^\alpha du\,\left( 1 - u^2\right)^{p-1} - \int_0^1 du\,\left( 1 - u^2\right)^{p-1} = -\int_\alpha^{1} du\,\left( 1 - u^2\right)^{p-1} = -\widetilde{U}_{(p)}(\alpha)
\end{equation}
where we recognized the definition of the integral \eqref{singint}. We get
\begin{equation}
-\lim_{\epsilon\rightarrow 0}\int_{\partial \mathcal{D}_\epsilon}\sqrt{H}\,\bigl( B_{(m)} - B_{(m)}\lvert_{\alpha=1} \bigr) = \Omega_{2p-1}\widetilde{C}_{(m,p)}\widetilde{U}_{(p)}(\alpha)\int_{A}\sqrt{h}\,\widehat{R}_{(m-p)}.
\end{equation}
This matches exactly with the expression \eqref{tildes} for $ D_{(m,p)}(\alpha) $.

For $ p>m $, the leading term \eqref{Bleading} with $ p-1 $ angular tensors $ \mathcal{Q}^{\phi_1\phi_2}_{\varphi_1\varphi_2} $ does not contribute to the sum. Hence all the terms are of order $ \mathcal{O}(\epsilon^{2}) $ so that
\begin{equation}
-\lim_{\epsilon\rightarrow 0}\int_{\partial \mathcal{D}_\epsilon}\sqrt{H}\,\bigl( B_{(m)} - B_{(m)}\lvert_{\alpha=1} \bigr) = 0.
\end{equation}
This is in agreement with the vanishing $ D_{(m,p)}(\alpha) $ for $ p>m $.

\section{Euclidean AdS\texorpdfstring{$ _{\mathbf{2m+1}} $}{1} in \texorpdfstring{$ S^{2p-1}\times \mathbb{H}^{2m-2p+1} $}{2}-slicing}\label{slicing}

In this Appendix, we describe a slicing of AdS$ _{2m+1} $ that appeared as the $ \alpha=1 $ limit of the hyperbolic brane geometry \eqref{hypsol}.

Consider the embedding of Euclidean AdS$ _{2m+1} $
\begin{equation}
-X_{0}^{2} + \sum_{i=1}^{2m+1}X_{i}^{2}=-\ell^{2}.
\end{equation}
into $ \mathbb{R}^{1,2m+1} $ with the metric
\begin{equation}
ds^{2} = -dX_0^{2}+\sum_{i=1}^{2m+1}dX_{i}^{2}.
\end{equation}
The embedding is solved by
\begin{align}
X_{0} &= r\cosh{u}\\
X_{i} &= 
\begin{cases}
(r^{2}-\ell^{2})^{1\slash 2}\,\cos{\theta}, \quad &i = 1\\
(r^{2}-\ell^{2})^{1\slash 2}\,\sin{\theta}\,\Omega_{i}, \quad &i = 2,\ldots,2p\\
r\sinh{u}\,\tilde{\Omega}_{i}, \quad &i = 2p+1,\ldots, 2m+1
\end{cases}
\end{align}
where
\begin{equation}
\sum_{i=2}^{2p}\Omega_{i}^{2} = 1, \quad \sum_{i=2p+1}^{2m+1}\tilde{\Omega}_{i}^{2} = 1.
\end{equation}
The ranges of the coordinates are
\begin{equation}
\ell\leq r, \quad  0 \leq u, \quad 0 \leq \theta < \pi.
\end{equation}
The resulting metric on AdS$ _{2m+1} $ is
\begin{equation}
ds^2 = (r^2 - \ell^2)\,(d\theta^{2} + \sin^{2}{\theta}\,d\Omega^2_{2p-2})  + \frac{\ell^2}{r^2-\ell^2}\,dr^2 + r^2( du^{2} + \sinh^{2}{u}\,d\tilde{\Omega}_{2m-2p}^2).
\label{hypslicing}
\end{equation}
This can also be written as
\begin{equation}
ds^2 = (r^2 - \ell^2)\, d\Omega^2_{2p-1} + \frac{\ell^2}{r^2-\ell^2}dr^2 + r^2d\Sigma_{2m-2p+1}^2
\end{equation}
which is the $ \alpha = 1 $ limit of the hyperbolic brane solution \eqref{hypsol} found in section \ref{hypbrane}. It corresponds to Euclidean AdS$ _{2m+1} $ in $ S^{2p-1}\times \mathbb{H}^{2m-2p+1} $-slicing. For $ p=1 $ the metric is Euclidean AdS--Rindler space which corresponds to $ S^{1}\times \mathbb{H}^{2m-1} $-slicing.

We compare the coordinates \eqref{hypslicing} to Poincaré coordinates. Poincar\'e coordinates give a flat slicing of AdS$ _{2m+1} $ and are obtained from
\begin{equation}
X_0 = \frac{\ell}{R}\frac{z}{2}\left[ 1 + \frac{R^{2} + \rho^{2} + \tilde{\rho}^{2}}{z^{2}}\right] , \quad X_1 = \frac{\ell}{R}\frac{z}{2}\left[ 1 - \frac{R^{2} - \rho^{2} - \tilde{\rho}^{2}}{z^{2}}\right] 
\end{equation}
\begin{equation}
X_{i} = 
\begin{cases}
\frac{\ell}{z}\rho\,\Omega_{i}, \quad &i = 2,\ldots,2p\\
\frac{\ell}{z}\tilde{\rho}\,\tilde{\Omega}_{i}, \quad &i = 2p+1,\ldots, 2m+1
\end{cases}
\end{equation}
where $ R $ is an arbitrary length scale and the coordinate ranges are
\begin{equation}
0\leq z,\rho,\tilde{\rho}.
\end{equation}
The resulting metric is
\begin{equation}
ds^{2} = \frac{\ell^{2}}{z^{2}}\left(dz^{2} +  d\rho^{2} + \rho^{2}d\Omega^{2}_{2p-2}+  d\tilde{\rho}^{2} + \tilde{\rho}^{2}d\tilde{\Omega}^{2}_{2m-2p} \right) 
\end{equation}
where the $ \mathbb{R}^{2m} $-slice has been factorized as $ \mathbb{R}^{2p-1}\times \mathbb{R}^{2m-2p+1} $ with each factor written in spherical coordinates.

The two coordinate systems are related by a transformation of the form
\begin{equation}
r=r(z,\rho,\tilde{\rho}), \quad \theta=\theta(z,\rho,\tilde{\rho}), \quad u=u(z,\rho,\tilde{\rho})
\end{equation}
with rest of the coordinates being the same between the two foliations. On the boundary $ r=\infty $ or $ z=0 $, the transformation induces the generalized Casini--Huerta--Myers conformal map \eqref{confmap}.

By equating $ X_1 $ coordinate of the two embeddings, we find that the surface $ r=\ell $ corresponds to the ($ 2m-2p+1 $)-dimensional hemisphere
\begin{equation}
z^{2} + \tilde{\rho}^{2} = R^{2}, \quad \rho = 0
\end{equation}
of radius $ R $ in Poincaré coordinates. On the boundary $ z=0 $, the hemisphere asymptotes to a sphere $ S^{2m-2p}_{R} $ of radius $ R $ embedded inside the $ \mathbb{R}^{2m-2p+1} $ factor of $ \mathbb{R}^{2m} $. This describes the relation between the spherical defect and the hyperbolic surface $ r=\ell $.

\end{appendix}

\printbibliography

\end{document}